\pdfoutput=1
\documentclass[12pt,a4paper]{article}
\usepackage{ifthen}
\newboolean{pdflatex}
\setboolean{pdflatex}{true}

\newboolean{articletitles}
\setboolean{articletitles}{true}

\newboolean{uprightparticles}
\setboolean{uprightparticles}{false}

\def\paperauthors{LHCb collaboration}
\def\paperasciititle{Searches for low-mass dimuon resonances}
\def\papertitle{Searches for \\ low-mass dimuon resonances}
\def\paperkeywords{{High Energy Physics}, {LHCb}}
\def\papercopyright{\the\year\ CERN for the benefit of the LHCb collaboration}
\def\paperlicence{CC BY 4.0 licence}
\def\paperlicenceurl{https://creativecommons.org/licenses/by/4.0/}

\def\xboson{\ensuremath{X}\xspace}
\def\xtomm{\ensuremath{\xboson\!\to\!\mu^+\mu^-}\xspace}
\def\mmm {\ensuremath{m(\mu^+\mu^-)}\xspace}
\def\mx{\ensuremath{m(\xboson)}\xspace}
\def\gx{\ensuremath{\Gamma(\xboson)}\xspace}
\def\tx{\ensuremath{\tau(\xboson)}\xspace}

\def\smm{\ensuremath{\sigma[\mmm]}\xspace}
\def\amm {\ensuremath{\alpha(\mu^+\mu^-)}\xspace}

\usepackage[top=1in, bottom=1.25in, left=1in, right=1in]{geometry}

\usepackage{microtype}
\usepackage{lineno}  
\usepackage{xspace} 
\usepackage{caption} 

\usepackage{graphicx}  
\usepackage{color}
\usepackage{colortbl}
\graphicspath{{./figs/}} 

\usepackage{amsmath} 
\usepackage{amssymb}
\usepackage{amsfonts}
\usepackage{upgreek} 

\newcommand*\patchAmsMathEnvironmentForLineno[1]{%
\expandafter\let\csname old#1\expandafter\endcsname\csname #1\endcsname
\expandafter\let\csname oldend#1\expandafter\endcsname\csname
end#1\endcsname
 \renewenvironment{#1}%
   {\linenomath\csname old#1\endcsname}%
   {\csname oldend#1\endcsname\endlinenomath}%
}
\newcommand*\patchBothAmsMathEnvironmentsForLineno[1]{%
  \patchAmsMathEnvironmentForLineno{#1}%
  \patchAmsMathEnvironmentForLineno{#1*}%
}
\AtBeginDocument{%
\patchBothAmsMathEnvironmentsForLineno{equation}%
\patchBothAmsMathEnvironmentsForLineno{align}%
\patchBothAmsMathEnvironmentsForLineno{flalign}%
\patchBothAmsMathEnvironmentsForLineno{alignat}%
\patchBothAmsMathEnvironmentsForLineno{gather}%
\patchBothAmsMathEnvironmentsForLineno{multline}%
\patchBothAmsMathEnvironmentsForLineno{eqnarray}%
}

\usepackage{hyperxmp}

\usepackage[pdftex,
            pdfauthor={\paperauthors},
            pdftitle={\paperasciititle},
            pdfkeywords={\paperkeywords},
            pdfcopyright={Copyright (C) \papercopyright},
            pdflicenseurl={\paperlicenceurl}]{hyperref}

\usepackage[bottom,flushmargin,hang,multiple]{footmisc}

\usepackage[all]{hypcap} 

\usepackage{xspace} 
\usepackage{upgreek}

\def\lhcb   {\mbox{LHCb}\xspace}

\def\MagUp {\mbox{\em Mag\kern -0.05em Up}\xspace}

\ifthenelse{\boolean{uprightparticles}}%
{

 \def\Ppsi        {\ensuremath{\uppsi}\xspace}

 \def\PDelta      {\ensuremath{\Delta}\xspace}                 
 \def\PXi         {\ensuremath{\Xi}\xspace}                 
 \def\PLambda     {\ensuremath{\Lambda}\xspace}                 
 \def\PSigma      {\ensuremath{\Sigma}\xspace}                 
 \def\POmega      {\ensuremath{\Omega}\xspace}                 
 \def\PUpsilon    {\ensuremath{\Upsilon}\xspace}

 \def\PB      {\ensuremath{\mathrm{B}}\xspace}                 
                  
 \def\PD      {\ensuremath{\mathrm{D}}\xspace}

 \def\PJ      {\ensuremath{\mathrm{J}}\xspace}                 
 \def\PK      {\ensuremath{\mathrm{K}}\xspace}

 \def\PZ      {\ensuremath{\mathrm{Z}}\xspace}                 
                  
 \def\Pb      {\ensuremath{\mathrm{b}}\xspace}                 
 \def\Pc      {\ensuremath{\mathrm{c}}\xspace}

 \def\Pi      {\ensuremath{\mathrm{i}}\xspace}

 \def\Ps      {\ensuremath{\mathrm{s}}\xspace}

 \def\thebaroffset{0.0em}
}
{

 \def\Ppsi        {\ensuremath{\psi}\xspace}                 
                  
 \mathchardef\PDelta="7101
 \mathchardef\PXi="7104
 \mathchardef\PLambda="7103
 \mathchardef\PSigma="7106
 \mathchardef\POmega="710A
 \mathchardef\PUpsilon="7107
                  
 \def\PB      {\ensuremath{B}\xspace}                 
                  
 \def\PD      {\ensuremath{D}\xspace}

 \def\PJ      {\ensuremath{J}\xspace}                 
 \def\PK      {\ensuremath{K}\xspace}

 \def\PZ      {\ensuremath{Z}\xspace}                 
                  
 \def\Pb      {\ensuremath{b}\xspace}                 
 \def\Pc      {\ensuremath{c}\xspace}

 \def\Pi      {\ensuremath{i}\xspace}

 \def\Ps      {\ensuremath{s}\xspace}

 \def\thebaroffset{0.18em}
}
\newcommand{\offsetoverline}[2][\thebaroffset]{\kern #1\overline{\kern -#1 #2}}%

\makeatletter
\ifcase \@ptsize \relax
  \newcommand{\miniscule}{\@setfontsize\miniscule{4}{5}}
\or
  \newcommand{\miniscule}{\@setfontsize\miniscule{5}{6}}
\or
  \newcommand{\miniscule}{\@setfontsize\miniscule{5}{6}}
\fi
\makeatother

\DeclareRobustCommand{\optbar}[1]{\shortstack{{\miniscule (\rule[.5ex]{1.25em}{.18mm})}
  \\ [-.7ex] $#1$}}

\def\Z      {{\ensuremath{\PZ}}\xspace}

\def\squark    {{\ensuremath{\Ps}}\xspace}

\def\cquark    {{\ensuremath{\Pc}}\xspace}

\def\bquark    {{\ensuremath{\Pb}}\xspace}

\def\kaon    {{\ensuremath{\PK}}\xspace}

\def\KorKbar {\kern \thebaroffset\optbar{\kern -\thebaroffset \PK}{}\xspace}

\def\KS      {{\ensuremath{\kaon^0_{\mathrm{S}}}}\xspace}

\def\D       {{\ensuremath{\PD}}\xspace}

\def\DorDbar {\kern \thebaroffset\optbar{\kern -\thebaroffset \PD}\xspace}

\def\Dp      {{\ensuremath{\D^+}}\xspace}
\def\Dm      {{\ensuremath{\D^-}}\xspace}

\def\DpDm    {\ensuremath{\Dp {\kern -0.16em \Dm}}\xspace}

\def\B       {{\ensuremath{\PB}}\xspace}

\def\BorBbar {\kern \thebaroffset\optbar{\kern -\thebaroffset \PB}\xspace}

\def\Bd      {{\ensuremath{\B^0}}\xspace}

\def\BdorBdbar {\kern \thebaroffset\optbar{\kern -\thebaroffset \Bd}\xspace}

\def\Bs      {{\ensuremath{\B^0_\squark}}\xspace}

\def\BsorBsbar {\kern \thebaroffset\optbar{\kern -\thebaroffset \Bs}\xspace}

\def\jpsi     {{\ensuremath{{\PJ\mskip -3mu/\mskip -2mu\Ppsi}}}\xspace}

\def\Upsilonres  {{\ensuremath{\PUpsilon}}\xspace}
\def\Y#1S{\ensuremath{\PUpsilon{(#1S)}}\xspace}

\def\LorLbar     {\kern \thebaroffset\optbar{\kern -\thebaroffset \PLambda}\xspace}

\def\to                 {\ensuremath{\rightarrow}\xspace}

\def\AT#1     {\ensuremath{A_{\mathrm{T}}^{#1}}\xspace}           

\def\C#1      {\ensuremath{\mathcal{C}_{#1}}\xspace}                       
\def\Cp#1     {\ensuremath{\mathcal{C}_{#1}^{'}}\xspace}                    
\def\Ceff#1   {\ensuremath{\mathcal{C}_{#1}^{\mathrm{(eff)}}}\xspace}        
\def\Cpeff#1  {\ensuremath{\mathcal{C}_{#1}^{'\mathrm{(eff)}}}\xspace}       
\def\Ope#1    {\ensuremath{\mathcal{O}_{#1}}\xspace}                       
\def\Opep#1   {\ensuremath{\mathcal{O}_{#1}^{'}}\xspace}                    

\newcommand{\nospaceunit}[1]{\ensuremath{\text{#1}}}       
\newcommand{\aunit}[1]{\ensuremath{\text{\,#1}}}       

\newcommand{\tev}{\aunit{Te\kern -0.1em V}\xspace}
\newcommand{\gev}{\aunit{Ge\kern -0.1em V}\xspace}
\newcommand{\mev}{\aunit{Me\kern -0.1em V}\xspace}
\newcommand{\kev}{\aunit{ke\kern -0.1em V}\xspace}
\newcommand{\ev}{\aunit{e\kern -0.1em V}\xspace}
 
\newcommand{\mevc}{\ensuremath{\aunit{Me\kern -0.1em V\!/}c}\xspace}
\newcommand{\gevc}{\ensuremath{\aunit{Ge\kern -0.1em V\!/}c}\xspace}
\newcommand{\mevcc}{\ensuremath{\aunit{Me\kern -0.1em V\!/}c^2}\xspace}
\newcommand{\gevcc}{\ensuremath{\aunit{Ge\kern -0.1em V\!/}c^2}\xspace}

\def\mm   {\aunit{mm}\xspace}

\def\mum  {\ensuremath{\,\upmu\nospaceunit{m}}\xspace}

\def\fb   {\ensuremath{\aunit{fb}}\xspace}
\def\invfb   {\ensuremath{\fb^{-1}}\xspace}

\def\gsim{{~\raise.15em\hbox{$>$}\kern-.85em
          \lower.35em\hbox{$\sim$}~}\xspace}
\def\lsim{{~\raise.15em\hbox{$<$}\kern-.85em
          \lower.35em\hbox{$\sim$}~}\xspace}

\def\pt         {\ensuremath{p_{\mathrm{T}}}\xspace}

\def\evtgen     {\mbox{\textsc{EvtGen}}\xspace}

\def\geant      {\mbox{\textsc{Geant4}}\xspace}

\def\photos     {\mbox{\textsc{Photos}}\xspace}

\def\pythia     {\mbox{\textsc{Pythia}}\xspace}

\def\tell1  {TELL1\xspace}
\def\ukl1   {UKL1\xspace}

\usepackage{cite} 
\usepackage{mciteplus}

\usepackage{longtable}
\usepackage{booktabs}
\usepackage{enumitem}

\begin{document}

\renewcommand{\thefootnote}{\fnsymbol{footnote}}
\setcounter{footnote}{1}

\begin{titlepage}
\pagenumbering{roman}

\vspace*{-1.5cm}
\centerline{\large EUROPEAN ORGANIZATION FOR NUCLEAR RESEARCH (CERN)}
\vspace*{1.5cm}
\noindent
\begin{tabular*}{\linewidth}{lc@{\extracolsep{\fill}}r@{\extracolsep{0pt}}}
\ifthenelse{\boolean{pdflatex}}
{\vspace*{-1.5cm}\mbox{\!\!\!\includegraphics[width=.14\textwidth]{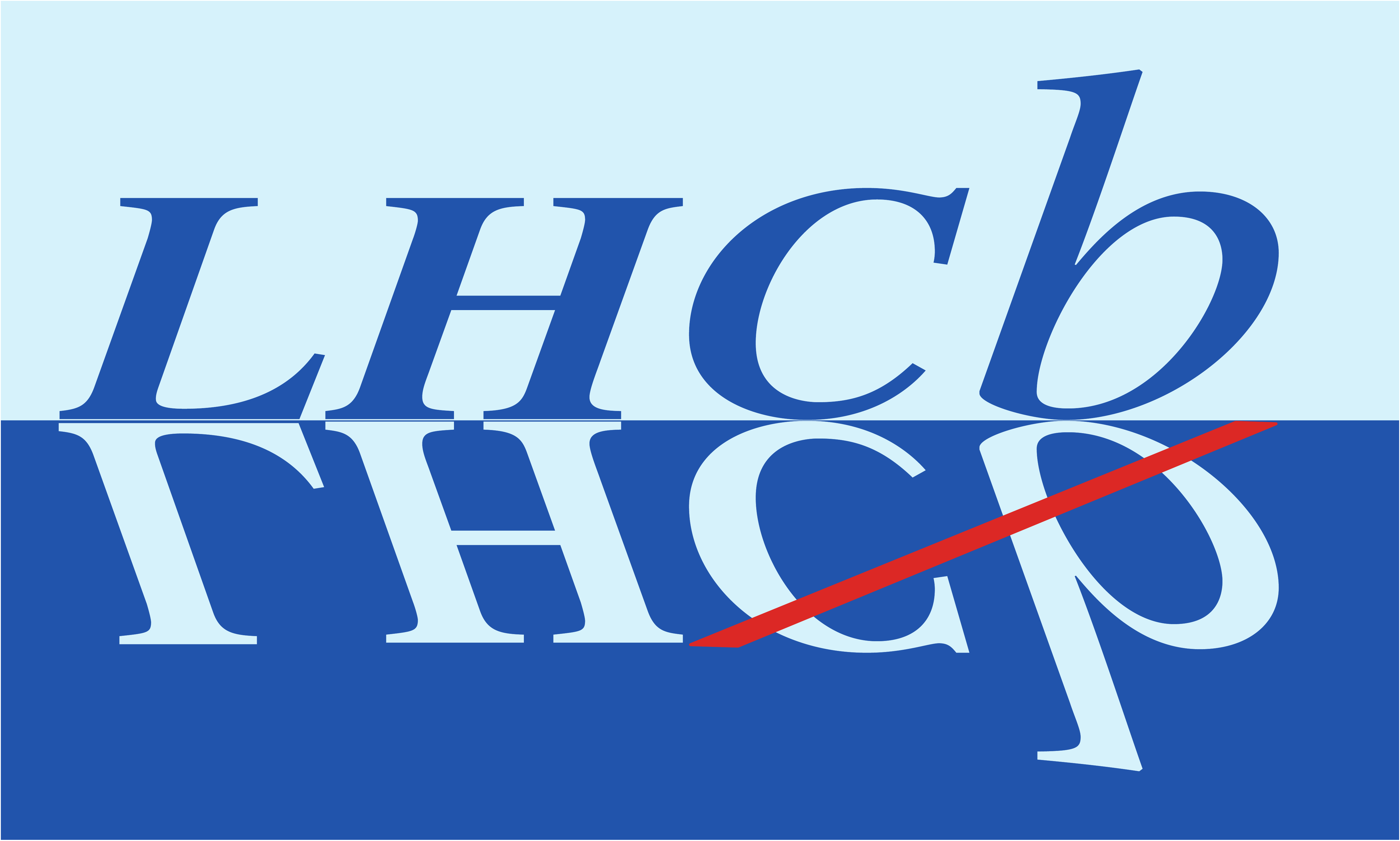}} & &}%
{\vspace*{-1.2cm}\mbox{\!\!\!\includegraphics[width=.12\textwidth]{lhcb-logo.eps}} & &}%
\\
 & & CERN-EP-2020-114 \\  
 & & LHCb-PAPER-2020-013 \\  
 & & November 2, 2020 \\ %
 & & \\
\end{tabular*}

\vspace*{3.0cm}

{\normalfont\bfseries\boldmath\huge
\begin{center}
  \papertitle
\end{center}
}

\vspace*{1.5cm}

\begin{center}
\paperauthors\footnote{Authors are listed at the end of this paper.}
\end{center}

\vspace{\fill}

\begin{abstract}
  \noindent
Searches are performed for a low-mass dimuon resonance, \xboson, produced in proton-proton collisions at a center-of-mass energy of 13\tev,
using a data sample corresponding to an integrated luminosity of 5.1\invfb and collected with the LHCb detector.
The \xboson bosons can either decay promptly or displaced from the proton-proton collision, where in both cases the requirements placed on the event and the assumptions made about the production mechanisms are kept as minimal as possible.
The searches for promptly decaying \xboson bosons explore the mass range from near the dimuon threshold up to 60\gev, with nonnegligible \xboson widths considered above 20\gev.
The searches for displaced \xtomm decays consider masses up to 3\gev.
None of the searches finds evidence for a signal and 90\% confidence-level exclusion limits are placed on the \xtomm cross sections, each with minimal model dependence.
In addition, these results are used to place world-leading constraints on \gev-scale bosons in the
two-Higgs-doublet and hidden-valley scenarios.

\end{abstract}

\vspace*{1.5cm}

\begin{center}
  Published in JHEP {\bf 10} (2020) 156
\end{center}

\vspace{\fill}

{\footnotesize
\centerline{\copyright~\papercopyright. \href{\paperlicenceurl}{\paperlicence}.}}
\vspace*{2mm}

\end{titlepage}

\newpage
\setcounter{page}{2}
\mbox{~}


\renewcommand{\thefootnote}{\arabic{footnote}}
\setcounter{footnote}{0}

\cleardoublepage


\pagestyle{plain}
\setcounter{page}{1}
\pagenumbering{arabic}


\section{Introduction}

Substantial effort has been dedicated~\cite{Battaglieri:2017aum}
to searching for a massive dark photon, $A'$, which obtains a small coupling to the electromagnetic current due to kinetic mixing between the Standard Model~(SM) hypercharge and $A'$ field strength tensors~\cite{Fayet:1980rr,Fayet:1980ad,Okun:1982xi,Galison:1983pa,Holdom:1985ag,Pospelov:2007mp,ArkaniHamed:2008qn,Bjorken:2009mm}.
However, this minimal $A'$ model is not the only viable dark-sector scenario.
The strongest connection to the dark sector may not arise via kinetic mixing, and the dark sector itself could be populated by additional particles that have phenomenological implications.
Searches for dark photons can provide serendipitous discovery potential for other types of particles, generically labeled here as \xboson bosons, especially vector particles that share the same production mechanisms as the minimal dark photon~\cite{Ilten:2018crw}, yet
many well-motivated types of \xboson bosons  would have avoided detection in all previous experimental searches~\cite{Fayet:1990wx,Berlin:2018tvf}.
For example, hidden-valley~(HV) scenarios that exhibit confinement produce a high multiplicity of light hidden hadrons from showering processes~\cite{Pierce:2017taw}.
These hidden hadrons would typically decay displaced from the proton-proton collision, thus failing the criteria employed in Refs.~\cite{LHCb-PAPER-2017-038,LHCb-PAPER-2019-031} to suppress backgrounds due to heavy-flavor quarks~\cite{Ilten:2015hya,Ilten:2016tkc}.
Furthermore, the sensitivity to various model scenarios can be improved
by exploiting additional signatures, {\em e.g.}, the presence of a \bquark-quark jet produced in association with the \xboson boson~\cite{Branco:2011iw}.
Therefore, it is desirable to perform searches that are less model dependent, including some that explore additional signatures in the event.

This article presents searches for low-mass dimuon resonances produced in proton-proton collisions at a center-of-mass energy of 13\tev,
using a data sample corresponding to an integrated luminosity of 5.1\invfb  and collected with the LHCb detector in 2016--2018.
The \xboson bosons can either decay promptly or displaced from the proton-proton collision.
In both cases, the requirements placed on the event and the assumptions made about the production mechanisms are  kept as minimal as possible.
Two variations of the search for prompt \xtomm decays are performed:
an inclusive version,
and an $X+b$ search, where the \xboson boson is required to be produced in association with a beauty quark.
Two variations are also considered of the search for displaced \xtomm decays:
an inclusive version,
and one where the \xboson boson is required to be produced promptly in the proton-proton collision.
The searches for prompt \xtomm decays explore the mass range from near the dimuon threshold up to 60\gev (natural units with $c=1$ are implied throughout this article), with nonnegligible widths, \gx, considered above 20\gev.
The searches for displaced \xtomm decays consider masses up to 3\gev.
This analysis uses the same data sample as the LHCb minimal dark-photon search~\cite{LHCb-PAPER-2019-031}; however, the searches presented here are roughly half as sensitive to the minimal $A'$ model, since the fiducial regions and selection criteria are not optimized for that scenario.
These searches are much more sensitive for many other \xboson boson scenarios, including HV models.

The fiducial regions used for each search, defined in Table~\ref{tab:fid},
ensure that the detector response is sufficiently model independent in the kinematic regions where results are reported.
The requirements placed on the momenta, $p$, and transverse momenta, \pt, of the muons make them sufficiently energetic to be selected by the trigger, but not so energetic that their charges cannot be determined.
Only events with at least one reconstructed proton-proton primary vertex (PV) are used in the analysis, which requires that at least five charged prompt particles, including the muons if the \xboson decays promptly, are produced in the same collision as the \xboson boson.
A maximum number of charged particles is allowed to be produced in the collision, since the detector response depends on the charged-particle multiplicity. In practice, this maximum value is sufficiently large to have no impact on any of the scenarios considered here.
The dimuon opening angle is required to be $\amm > 1\, (3)$\,mrad in the searches for prompt (displaced) \xtomm decays to ensure that the reconstruction efficiency factorizes into the product of the two individual muon efficiencies, which subsequently leads to an upper limit on $\pt(X)$ to remove regions where the \amm requirement is rarely satisfied.
The $X+b$ analysis is performed using jets clustered with the anti-$k_{\rm T}$ algorithm~\cite{antikt} using a distance parameter $R = 0.5$. The jets are required to have $20 < \pt({\rm jet}) < 100\gev$ and a pseudorapidity in the range $2.2 < \eta({\rm jet}) < 4.2$ so that the $b$-tagging efficiency is nearly uniform within the fiducial region.
Finally, the displaced \xtomm secondary vertex (SV) is required to be transversely displaced from the PV in the range $12 < \rho_{\rm T} < 30\mm$, which results in minimal dependence on the SV location distribution. For example, this requirement leads to the efficiency being nearly independent of the \xboson lifetime, \tx; however, the probability that the \xboson boson decays in this region is strongly dependent on \tx.

\begin{table}[t!]
    \caption{\label{tab:fid} Fiducial regions of the searches for prompt and displaced \xtomm decays.}\vspace{-0.3cm}
    \begin{center}
      \begin{tabular}{cc}
        \toprule
        {} &
         $\pt(\mu) > 0.5\gev$ \\ {}  & $10 \!<\! p(\mu) \!<\! 1000\gev$ \\
        All searches & $2 \!<\! \eta(\mu) \!<\! 4.5$ \\
         {} & $\sqrt{\pt(\mu^+) \pt(\mu^-)} > 1\gev$ \\
        {} & $5 \leq n_{\rm charged}(2 \!<\! \eta \!<\! 4.5, p\! >\! 5\gev)\! <\! 100$ (from same PV as \xboson) \\
        \midrule
        {}  &
         $ 1 < \pt(X) < 50\gev$ \\
         Prompt & $X$ decay time $< 0.1$\,ps \\
        \xtomm decays & $\amm > 1$\,mrad \\
         {} & $20 < \pt(b$-jet$) < 100\gev$,\, $2.2 < \eta(b$-jet$) < 4.2$ ($X+b$ only)\\
        \midrule
        {} &
         $2 < \pt(X) < 10\gev$ \\
         Displaced  & $2 \!<\! \eta(X) \!<\! 4.5$ \\
        \xtomm decays & $\amm >3$\,mrad \\
        {} & $12 < \rho_{\rm T}(X) < 30\mm$  \\
        {} & $X$ produced in $pp$ collision (promptly produced $X$ only) \\
        \bottomrule
      \end{tabular}
  \end{center}
\end{table}

This article is structured as follows.
The LHCb detector, trigger, and simulation are described in Sec.~\ref{sec:lhcb}, while the offline selections used in each of the searches are discussed in Sec.~\ref{sec:sel}.
Section~\ref{sec:search} presents the searches for  both prompt and displaced \xtomm decays.
Section~\ref{sec:eff} discusses the efficiencies and luminosity.
The model-independent cross-section results, along with their interpretations within the context of specific models, are described in Sec.~\ref{sec:results}.
Section~\ref{sec:sum} provides a summary and discussion of all results.

\section{Detector and simulation}
\label{sec:lhcb}

The \lhcb detector~\cite{LHCb-DP-2008-001,LHCb-DP-2014-002} is a single-arm forward
spectrometer covering the \mbox{pseudorapidity} range $2<\eta <5$,
designed for the study of particles containing \bquark or \cquark
quarks. The detector includes a high-precision tracking system
consisting of a silicon-strip vertex detector surrounding the proton-proton
interaction region (VELO), a large-area silicon-strip detector located
upstream of a dipole magnet with a bending power of about
$4{\mathrm{\,Tm}}$, and three stations of silicon-strip detectors and straw
drift tubes
placed downstream of the magnet.
The tracking system provides a measurement of the momentum of charged particles with
a relative uncertainty that varies from 0.5\% at low momentum to 1.0\% at 200\gev.
The minimum distance of a track to a PV, the impact parameter,
is measured with a resolution of $(15+29/\pt)\mum$,
where \pt is
in\,\gev.
Different types of charged hadrons are distinguished using information
from two ring-imaging Cherenkov detectors.
Photons, electrons and hadrons are identified by a calorimeter system consisting of
scintillating-pad and preshower detectors, an electromagnetic
and a hadronic calorimeter. Muons are identified by a
system composed of alternating layers of iron and multiwire
proportional chambers.

The online event selection is performed by a trigger, which consists of a hardware stage followed by a two-level software stage.
In between the two software stages, an alignment and calibration of the detector is performed in near real-time and their results are used in the trigger~\cite{LHCb-PROC-2015-011}.
The same alignment and calibration information is propagated
to the offline reconstruction, ensuring consistent and high-quality
particle identification information between the trigger and
offline software. The identical performance of the online and offline
reconstruction offers the opportunity to perform physics analyses
directly using candidates reconstructed in the trigger
\cite{LHCb-DP-2012-004,LHCb-DP-2016-001},
which
the searches for prompt \xtomm decays
exploit.

At the hardware trigger stage, events are required to have a dimuon pair with $\pt(\mu^+) \pt(\mu^-) \gtrsim (1.5\gev)^2$
and at most 900 hits in the scintillating-pad detector, which prevents high-occupancy events from dominating the processing time in the software trigger stages.
The latter requirement is the main motivation for defining the maximum charged-particle multiplicity in Table~\ref{tab:fid}.
In the software stage, where the \pt resolution is substantially improved compared to the hardware stage,
\xtomm candidates are built from two oppositely charged tracks
that form a good-quality vertex and satisfy stringent muon-identification criteria.
All searches require $\pt(\xboson) > 1\gev$ and $2<\eta(\mu)<4.5$.
The searches for prompt \xtomm decays use muons that are consistent with originating from the PV,
with ${\pt(\mu) > 1.0\gev}$ and momentum ${p(\mu) > 20\gev}$ in the 2016 data sample,  and $\pt(\mu) > 0.5\gev$, ${p(\mu) > 10\gev}$, and ${\pt(\mu^+) \pt(\mu^-) > (1.0\gev)^2}$ in \mbox{2017--2018}.
The searches for displaced \xtomm decays use muons with $\pt(\mu) > 0.5\gev$ and ${p(\mu) > 10\gev}$
that are inconsistent with originating from any PV,
and require $2 < \eta(\xboson)<4.5$.  In addition, the search for a long-lived  promptly produced \xboson boson requires a decay topology consistent with a dimuon resonance originating from a PV.

Simulation is required to model the effects of the detector acceptance and its response to \xtomm decays.
In the simulation, $pp$ collisions are generated using
\pythia~\cite{Sjostrand:2007gs,*Sjostrand:2006za}
with a specific \lhcb configuration~\cite{LHCb-PROC-2010-056}.
Decays of unstable particles
are described by \evtgen~\cite{Lange:2001uf}, in which final-state
radiation is generated using \photos~\cite{Golonka:2005pn}.
The interaction of the generated particles with the detector, and its response,
are implemented using the \geant
toolkit~\cite{Allison:2006ve, *Agostinelli:2002hh} as described in
Ref.~\cite{LHCb-PROC-2011-006}.
Simulation is also used to place constraints on specific models. Prompt limits for light-pseudoscalar models are set with next-to-next-to-leading order cross-sections from \textsc{Higlu}~\cite{Spira:1995rr,*Spira:1995mt} using the \textsc{Nnpdf}3.0 PDF set~\cite{Ball:2014uwa}, branching fractions from \textsc{Hdecay}~\cite{Djouadi:1997yw,*Djouadi:2018xqq}, and fiducial acceptances from \pythia~\cite{Sjostrand:2014zea}. Displaced limits for HV models are set with \pythia~\cite{Sjostrand:2014zea} using a running $\alpha_{\text{HV}}$ scheme~\cite{Schwaller:2015gea}, and couplings from \textsc{Darkcast}~\cite{Ilten:2018crw}.

\section{Selection}
\label{sec:sel}

\begin{figure}[t!]
  \centering
  \includegraphics[width=0.9\textwidth]{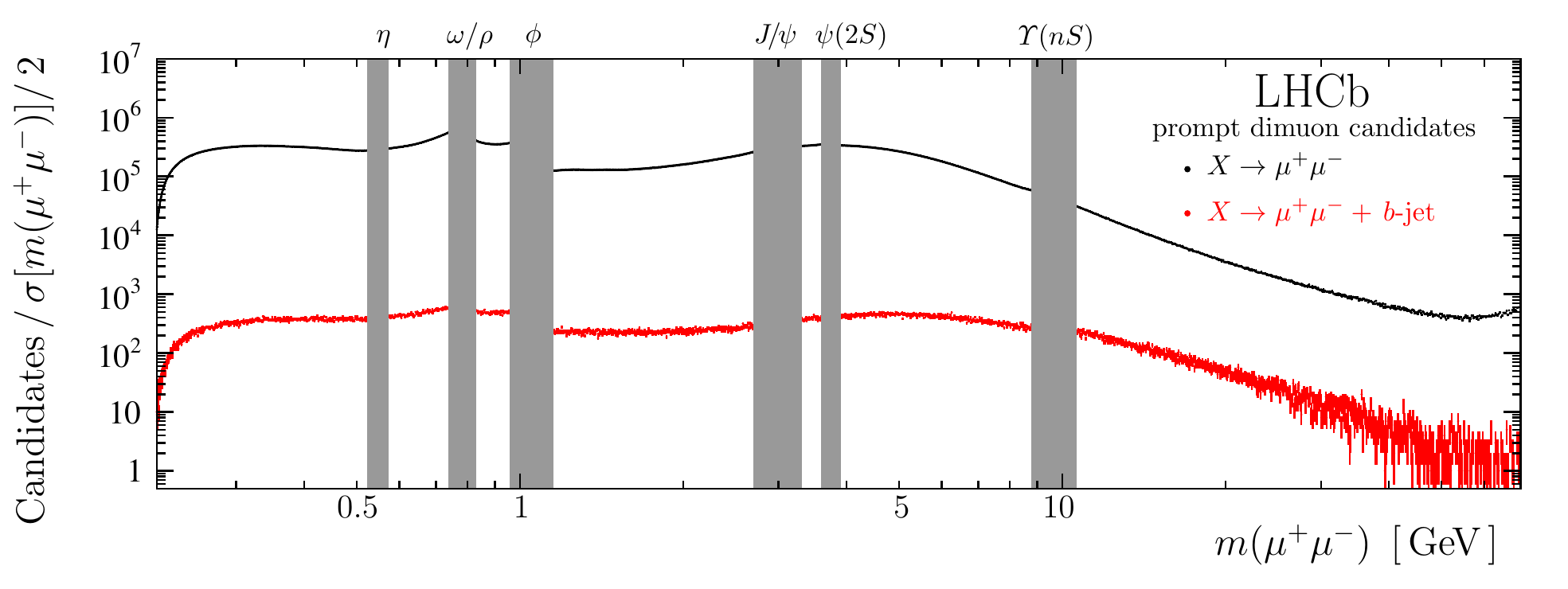}\vspace{-0.2cm}
  \caption{
  Prompt dimuon mass spectra showing the (black) inclusive and (red) $X+b$  candidates with all fiducial and selection requirements applied. The grey boxes show the regions vetoed due to large contributions from QCD resonances.
  }
  \label{fig:mprompt}
\end{figure}

The selection criteria are largely applied online in the trigger and most are the same as those used in the LHCb minimal dark-photon search~\cite{LHCb-PAPER-2019-031}.
The prompt dimuon sample, {\em i.e.}\ the sample used in the searches for prompt \xtomm decays, selected by the trigger described in Sec.~\ref{sec:lhcb} predominantly consists of genuine prompt dimuon pairs.
The only selection criteria applied offline in the inclusive search for prompt \xtomm decays, $\pt(X) < 50\gev$ and $\amm > 1$\,mrad, are included in the definition of the fiducial region.
In addition to these, the search for a promptly decaying \xboson boson produced in association with a beauty quark requires at least one $b$-tagged jet with $\pt({\rm jet}) > 20\gev$ and $2.2 < \eta({\rm jet}) < 4.2$.
The jets are formed by clustering charged and neutral particle-flow candidates~\cite{LHCb-PAPER-2013-058} using the anti-$k_{\rm T}$ clustering algorithm as implemented in {\sc FastJet}~\cite{fastjet}.
The $b$-tagging requires an SV in the jet that satisfies the criteria given in Ref.~\cite{LHCb-PAPER-2015-016}.
Figure~\ref{fig:mprompt} shows the \mmm distributions of both prompt dimuon data samples in bins of width $\smm / 2$, where \smm denotes the dimuon invariant-mass resolution which varies from 0.6\mev near threshold to 0.6\gev at $\mmm = 60\gev$.

\begin{figure}[t!]
  \centering
  \includegraphics[width=0.9\textwidth]{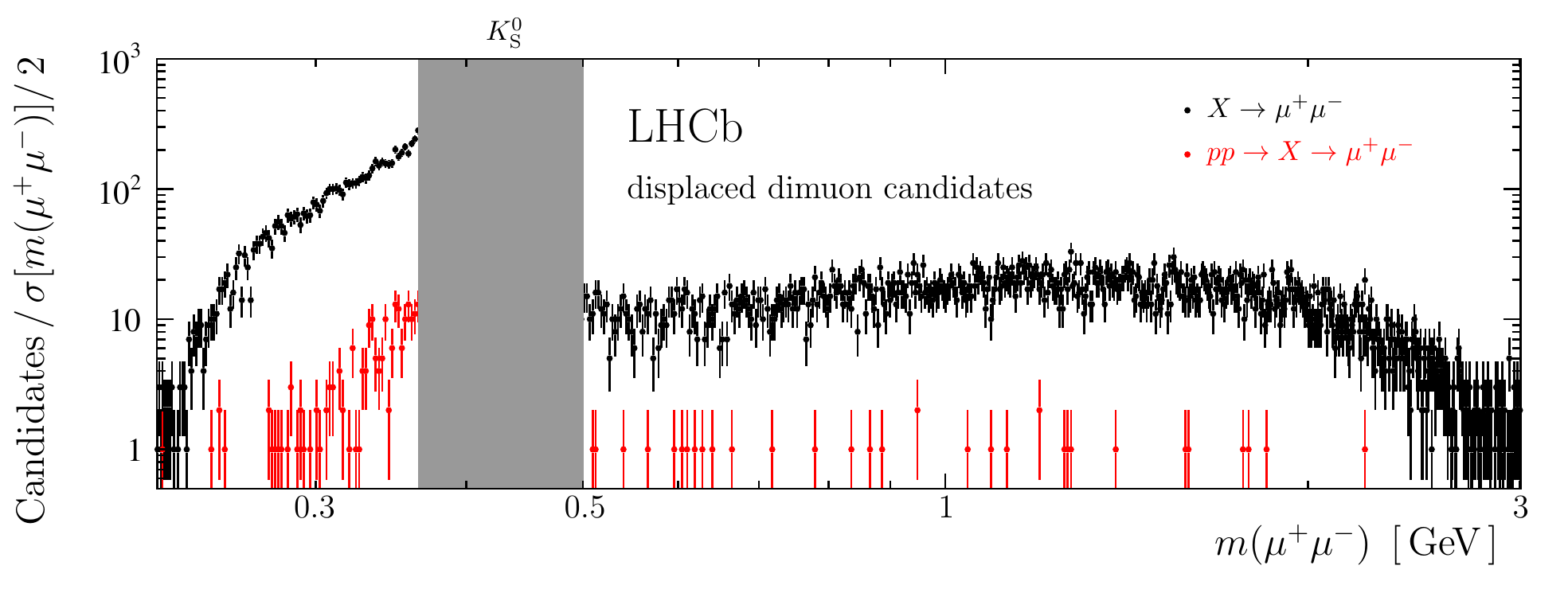}\vspace{-0.2cm}
  \caption{
  Displaced dimuon mass spectra  showing the (black) inclusive and (red) promptly produced candidates with all fiducial and selection requirements applied. The grey box shows the region vetoed due to the large doubly misidentified \KS background, whose low-mass tail  extends into the search region. A dedicated study of this region is presented in a search for $\KS \to \mu^+\mu^-$ decays~\cite{LHCb-PAPER-2017-009}.
  }
  \label{fig:mdispl}
\end{figure}

In the searches for displaced \xtomm decays, contamination from prompt particles is negligible due to a stringent trigger criterion that requires muons to be inconsistent with originating from any PV.
Furthermore, the fiducial region requires a transverse displacement from the PV of $12 < \rho_{\rm T} < 30\mm$, which is applied offline in both searches for displaced \xtomm decays and highly suppresses the background from $b$-hadron decay chains that produce two muons.
Therefore, the dominant background contributions are due to material interactions in the VELO, {\em e.g.}\ photons that convert into $\mu^+\mu^-$ pairs, and from $\KS\to\pi^+\pi^-$ decays, where both pions are misidentified as muons.
A $p$-value is assigned to the material-interaction hypothesis for each displaced \xtomm candidate using properties of the SV and muon tracks, along with a high-precision three-dimensional material map produced from a data sample of secondary hadronic interactions~\cite{LHCb-DP-2018-002}.
The same mass-dependent requirement used in Ref.~\cite{LHCb-PAPER-2019-031} is applied to the $p$-values in this analysis, which highly suppresses the material-interaction background.
Figure~\ref{fig:mdispl} shows the \mmm distributions of both displaced-dimuon data samples.

\section{Signal searches}
\label{sec:search}

The signal-search strategies and methods employed are similar to those used in Ref.~\cite{LHCb-PAPER-2019-031}.
The dimuon mass spectra are scanned in around $6000$ steps of about $\smm / 2$ searching for \xtomm contributions.
For $\mx < 20\gev$, the data are binned in $\pt(X)$ and each \pt bin is searched independently for each \mx hypothesis; whereas at higher masses, \pt bins are not necessary since both the resolution and efficiency are nearly independent of $\pt(X)$.
All searches use the profile likelihood method to determine the local $p$-values and the confidence intervals on the signal yields.
The trial factors are obtained using pseudoexperiments in each search.
The confidence intervals are defined using the {\em bounded likelihood} approach~\cite{Rolke:2004mj},
which involves taking $\Delta \log{\mathcal{L}}$ relative to zero signal, rather than the best-fit value, if the best-fit signal value is negative.
This approach enforces that only physical (nonnegative) upper limits are placed on the signal yields, and prevents defining exclusion regions that are much better than the experimental sensitivity in cases where a large deficit in the background yield is observed.
The signal \mmm distributions are well modeled by a Gaussian function, whose resolution is determined with 10\% precision using a combination of simulated \xtomm decays and the observed \pt-dependent widths of the large known resonance peaks  present in the data. The mass-resolution uncertainty is included in the profile likelihood.
The potential bias due to neglecting non-Gaussian components of the signal shape is much smaller than the uncertainty that arises from the limited knowledge of the mass resolution.

The fit strategy used in the searches for prompt \xtomm decays below 20\gev, which is the same as in Refs.~\cite{LHCb-PAPER-2017-038,LHCb-PAPER-2019-031},
was first introduced in Ref.~\cite{Williams:2017gwf}.
At each \mx hypothesis, a binned extended maximum-likelihood fit is performed in a $\pm12.5\,\smm$ window around the \mx value.
Near the dimuon threshold, the energy released in the decay, $Q = \sqrt{\mmm^2 - 4m(\mu)^2}$, is used instead of the mass because it is easier to model.
The background model for each fit window takes as input a large set of potential components, then the data-driven model-selection process of Ref.~\cite{Williams:2017gwf} is performed, whose uncertainty is included in the profile likelihood following Ref.~\cite{Dauncey:2014xga}.
Specifically, the method labeled {\em aic-o} in Ref.~\cite{Williams:2017gwf} is used, where the log-likelihood of each background model is penalized for its complexity (number of parameters).
The confidence intervals are obtained from the profile likelihoods, including the penalty terms, where the model index is treated as a discrete nuisance parameter, as originally proposed in Ref.~\cite{Dauncey:2014xga}.
In the $X+b$ search there are not many candidates near the dimuon threshold. Therefore, just in this region, the counting-experiment-based method of Ref.~\cite{Williams:2015xfa} is used, which is also used in the searches for displaced \xtomm decays and described in detail below.

In this analysis, the set of possible background components is the same as in Ref.~\cite{LHCb-PAPER-2019-031} and includes all Legendre modes up to tenth order at every \mx.
Additionally, dedicated background components are included for sizable narrow SM resonance contributions.
The use of 11 Legendre modes adequately describes every doubly misidentified peaking background that contributes at a significant level; therefore, these do not require dedicated background components.
In mass regions where such complexity is not required, the data-driven model-selection procedure reduces the complexity, which increases the sensitivity to a potential signal contribution.
Therefore, the impact of the background-model uncertainty on the size of the confidence intervals is mass dependent, though on average it is about 30\%.
As in Ref.~\cite{Williams:2017gwf}, all fit regions are transformed onto the interval $[-1,1]$, where the \mx value is mapped to zero.
After such a transformation, the signal model is (approximately) an even function;
therefore, odd Legendre modes are orthogonal to the signal component, which means that the presence of odd modes has minimal impact on the variance of the observed signal yield.
In the fits, all odd Legendre modes up to ninth order are included in every background model, while even modes must be selected for inclusion in each fit by the data-driven method of Ref.~\cite{Williams:2017gwf}.

Regions in the mass spectrum with large SM resonance contributions
are vetoed in the searches for prompt \xtomm decays.
Furthermore, the region near the $\eta^{\prime}$ meson
is treated uniquely.
Since it is not possible to distinguish between \xtomm and possible $\eta^{\prime}\!\to\!\mu^+\mu^-$ contributions at $m(\eta^{\prime})$, the $p$-values near this mass are ignored.
The small observed excess at $m(\eta^{\prime})$ is simply absorbed into the signal yield when setting the limits, which is conservative in that the $\eta^{\prime}\!\to\!\mu^+\mu^-$ contribution weakens the constraints on \xtomm decays.

\begin{figure}[t!]
  \centering
  \includegraphics[width=0.9\textwidth]{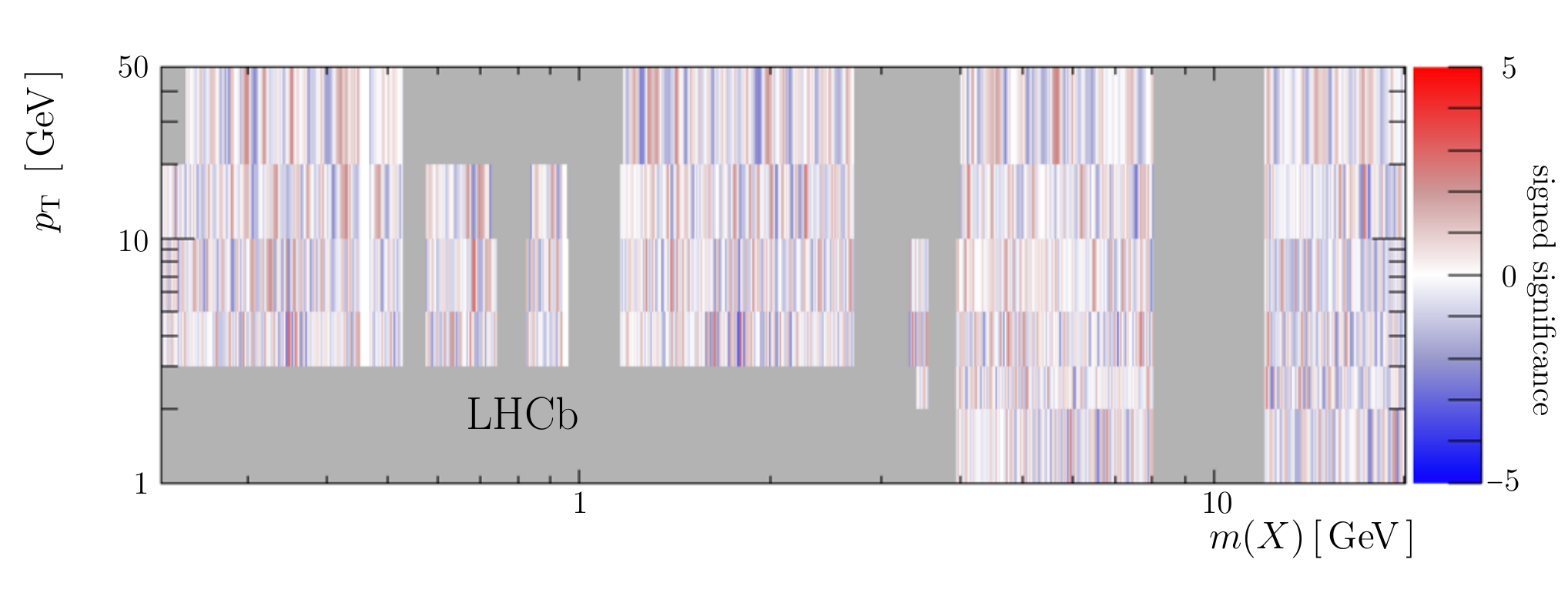}\\\vspace{-0.2cm}
   \includegraphics[width=0.9\textwidth]{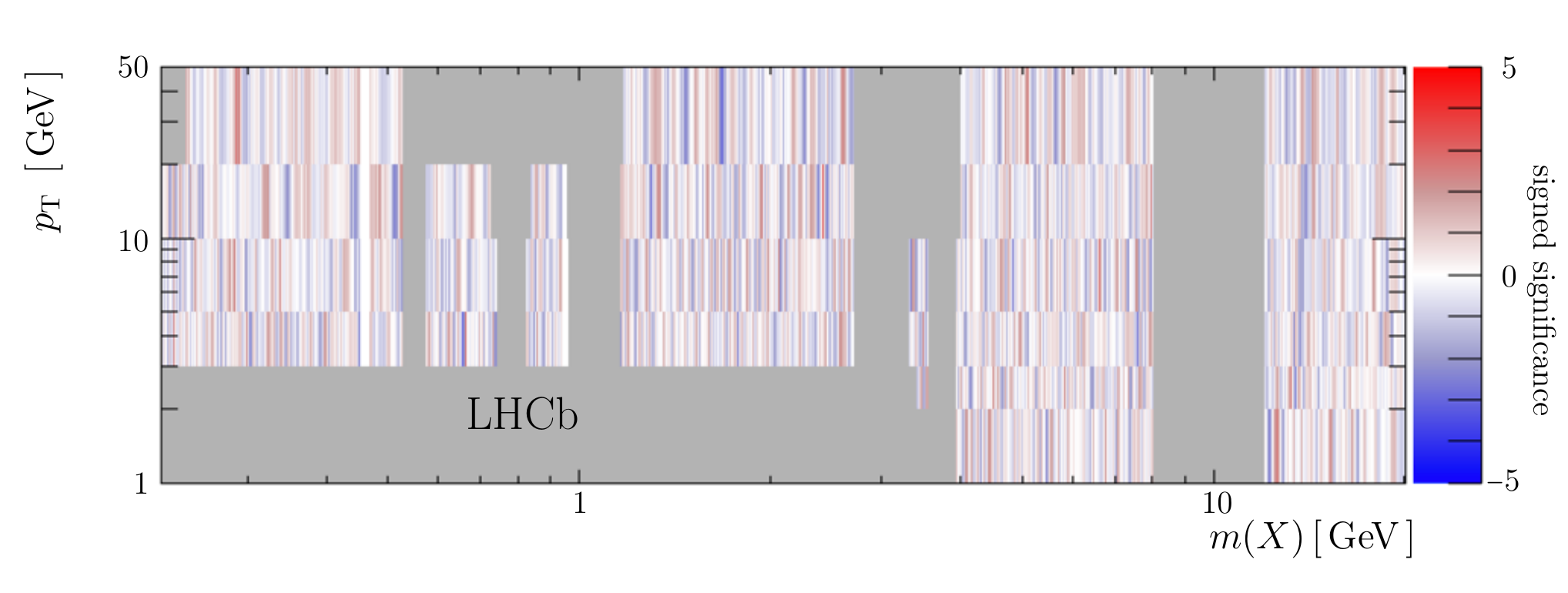}\vspace{-0.2cm}
  \caption{
  Signed local significances in the $\mx < 20\gev$ region for the (top) inclusive and (bottom) associated beauty searches for prompt \xtomm decays. If the best-fit signal-yield estimator is negative, the signed significance is negative and {\em vice versa}.
  The grey regions are excluded either due to a nearby large QCD resonance contribution, or because the overlap of the bin with the fiducial region in Table~\ref{tab:fid} is small.
  }
  \label{fig:prompt-sgnf}
\end{figure}

Figure~\ref{fig:prompt-sgnf} shows the signed local significances for all \mx below $20\gev$ for both searches for prompt \xtomm decays.
The largest local excess in the inclusive search in this mass region is $3.7\sigma$ at $349\mev$ in the $3 < \pt(X) < 5\gev$ bin; however, its neighboring \pt bin at this mass has a small deficit and the global significance is only $\approx 1\sigma$.
Similarly, the largest local excess in the $X+b$ search below 20\gev is $3.1\sigma$ at $2424\mev$ in the $10 < \pt(X) < 20\gev$ bin, though again, the neighboring \pt bins both have deficits at the same mass, and the global significance is below $1\sigma$.
Therefore, no significant excess is found in either prompt spectrum for $\mx < 20\gev$.

In the $20 < \mx < 60\gev$ region, the background is nearly monotonic, which permits the use of a simplified fit strategy.
The entire  $12 < \mmm < 80\gev$ region is fitted when considering all \mx values above 20\gev.
The background model is comprised of three falling power-law terms and an
eighth-order polynomial that collectively describe the Drell--Yan, heavy-flavor, and misidentified-background contributions, along with a rising power-law term to describe the low-mass tail of the \Z boson, where all parameters are free to vary.
This background model is validated by studying simulated Drell--Yan dimuon production, same-sign dimuon data which predominantly consists of heavy-flavor and misidentification backgrounds, and candidates in the data sample itself above the search region.
Unlike at lower masses, nonnegligible widths are considered. At each \mx, a scan is performed covering the range $0 \leq \gx \leq 3\gev$.
The signals are modelled by a Gaussian resolution function convolved with the modulus of a Breit--Wigner function.

\begin{figure}[t!]
  \centering
  \includegraphics[width=0.9\textwidth]{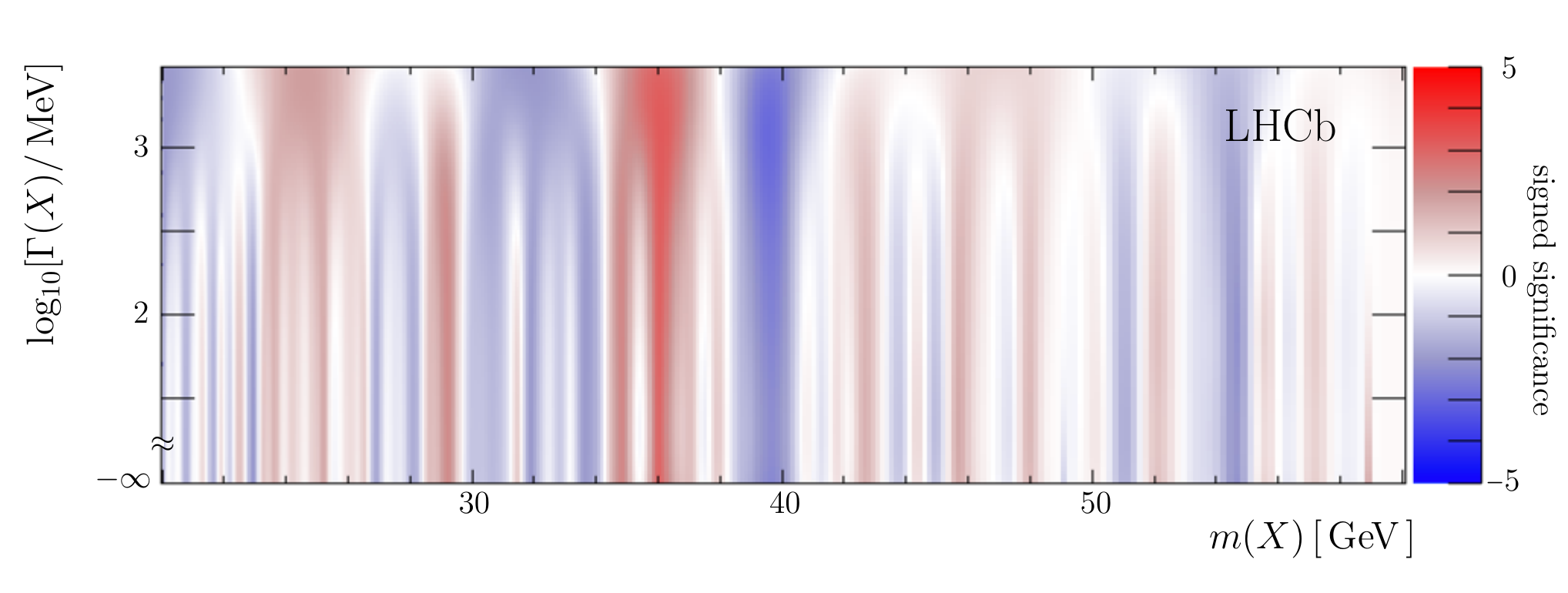}\\\vspace{-0.2cm}
  \includegraphics[width=0.9\textwidth]{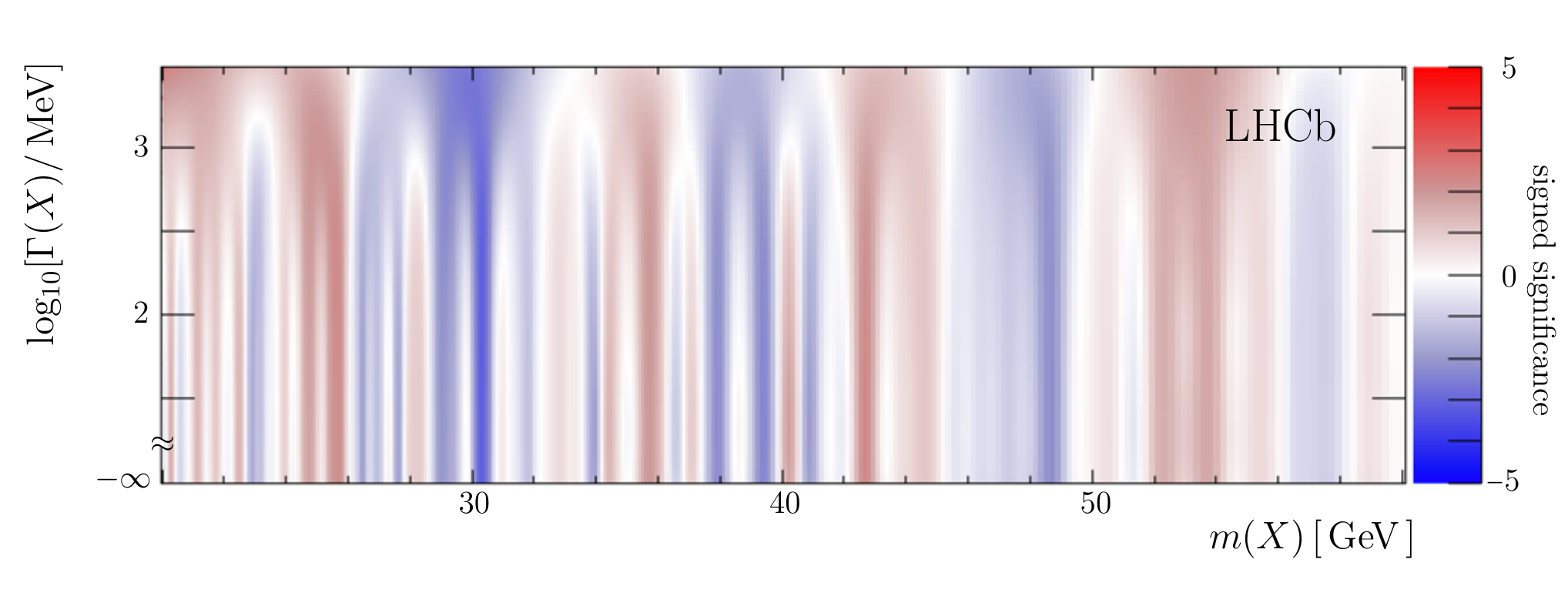}\vspace{-0.2cm}
  \caption{
  Signed local significances in the $\mx > 20\gev$ region for the (top) inclusive and (bottom) associated beauty searches for prompt \xtomm decays. The lower limit on the vertical axis of $\log_{10}[\gx / \mev] = -\infty$ corresponds to $\gx = 0$.
  }
  \label{fig:prompt-sgnf-highmass}
\end{figure}

Figure~\ref{fig:prompt-sgnf-highmass} shows the signed local significances for the $\mx > 20\gev$ region for both searches for prompt \xtomm decays.
The largest local excess in the inclusive search in this mass region is $3.2\sigma$ at $\mx = 36\gev$ for $\gx = 1.5\gev$, which corresponds to a global $p$-value of about 11\% (considering only the $\mx > 20\gev$ mass region).
In the $X+b$ search, no local significance exceeds $\approx 2\sigma$ in this mass region.
Therefore, no significant excess is found in either prompt spectrum for $\mx > 20\gev$.

\begin{figure}[t!]
  \centering
  \includegraphics[width=0.9\textwidth]{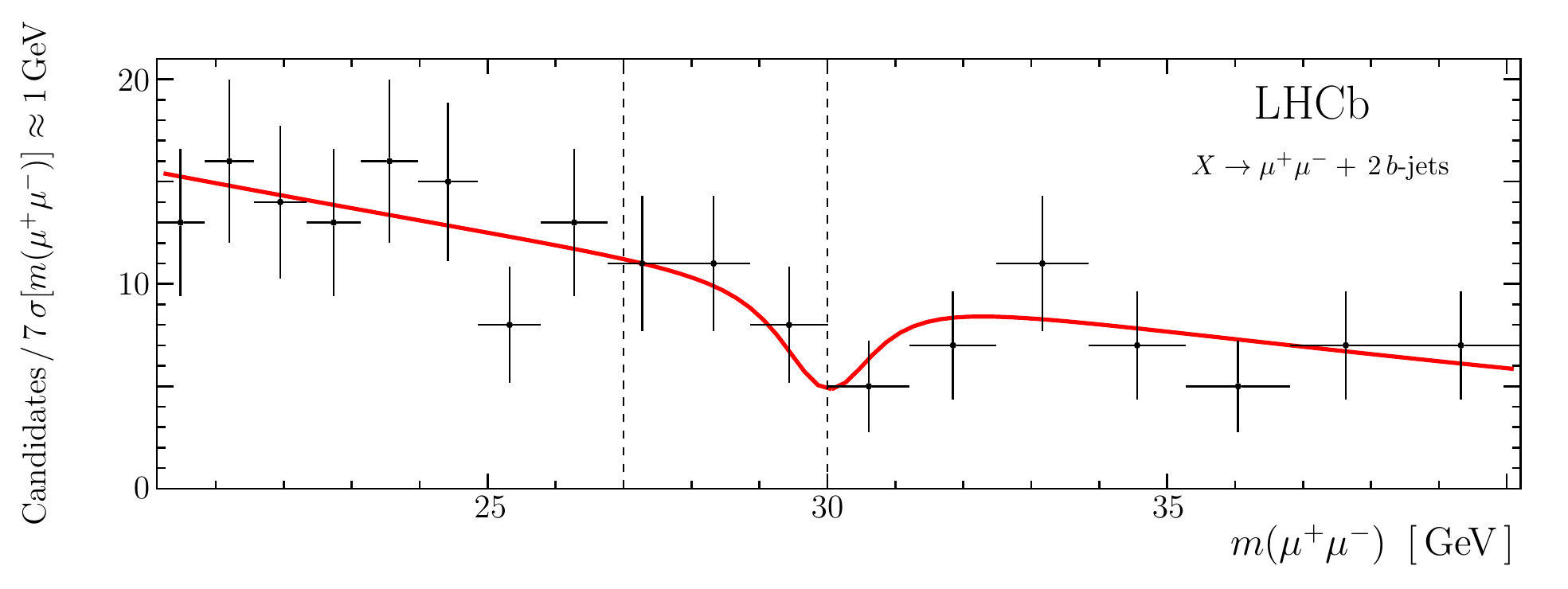}\vspace{-0.2cm}
  \caption{
  Fit to the \mmm spectrum in events with at least two $b$-tagged jets.
  The ${27 < \mx < 30\gev}$ search region is marked by the vertical dashed lines.
  }
  \label{fig:2bjet}
\end{figure}

Motivated by the possible excess seen by CMS~\cite{Sirunyan:2018wim} in $X+b\bar{b}$ events, a dedicated search for a resonance with $27 < \mx < 30\gev$ and $0.5 < \gx < 3.0\gev$ is performed in the subset of the $X+b$ candidates that contains at least two $b$-tagged jets.
The mass spectrum in the range 20--40\gev is fitted using a model consisting of a second-order polynomial background and a signal whose mass and width are free to vary within the \mx and \gx ranges specified above.
Figure~\ref{fig:2bjet} shows the result of this fit.
The best-fit signal yield is negative in the region considered; therefore, no evidence for a signal is observed.
Using the efficiency and luminosity from Sec.~\ref{sec:eff}, and their associated uncertainties, the upper limits on the $X(\mu^+\mu^-)+b\bar{b}$ cross section in the \mx and \gx regions considered are no larger than $15 \fb \times \sqrt{\gx / \gev}$.

The fit strategy used in the searches for displaced \xtomm decays below the \KS mass is also the same as in Refs.~\cite{LHCb-PAPER-2017-038,LHCb-PAPER-2019-031}.
Binned extended maximum-likelihood fits are performed to the $Q$ spectrum in each \pt bin.
The region near the \KS mass is vetoed to avoid the sizable background from doubly misidentified $\KS \to \pi^+\pi^-$ decays.
The expected photon-conversion contribution is derived from a sample of candidates that are consistent with a photon originating from a PV.
Two large control  samples are used to develop and validate the modeling of the \KS and remaining material-interaction contributions:
dimuon candidates that fail, but nearly satisfy, the stringent muon-identification criteria;
and a sample of dimuon candidates that is rejected by the material-interaction criterion.
Both contributions are well modeled by second-order polynomials in $Q$  below the \KS veto region.
The material-interaction contribution, apart from the dedicated photon-conversion component, is not needed in the search that requires a decay topology consistent with an \xboson boson originating from a PV.

The fit strategy used in the searches for displaced \xtomm decays above the \KS veto region, specifically, in the $0.5 < \mx < 3.0\gev$ mass range, is the same as used in the LHCb search for hidden-sector bosons produced in $\Bd \to K^{(*)} X(\mu^+\mu^-)$ decays~\cite{LHCb-PAPER-2015-036,LHCb-PAPER-2016-052}.
This strategy was first introduced in Ref.~\cite{Williams:2015xfa}.
Since no sharp features are expected in the background in this region, and due to the small bin occupancies, the background is estimated by interpolating the yields in the sidebands starting at $\pm3\smm$ from \mx.
The statistical test at each mass is based on the profile likelihood ratio of Poisson-process hypotheses with and without a signal contribution.
The uncertainty on the linearity of the background interpolation is modeled by a Gaussian term in the likelihood.

\begin{figure}[t!]
  \centering
  \includegraphics[width=0.9\textwidth]{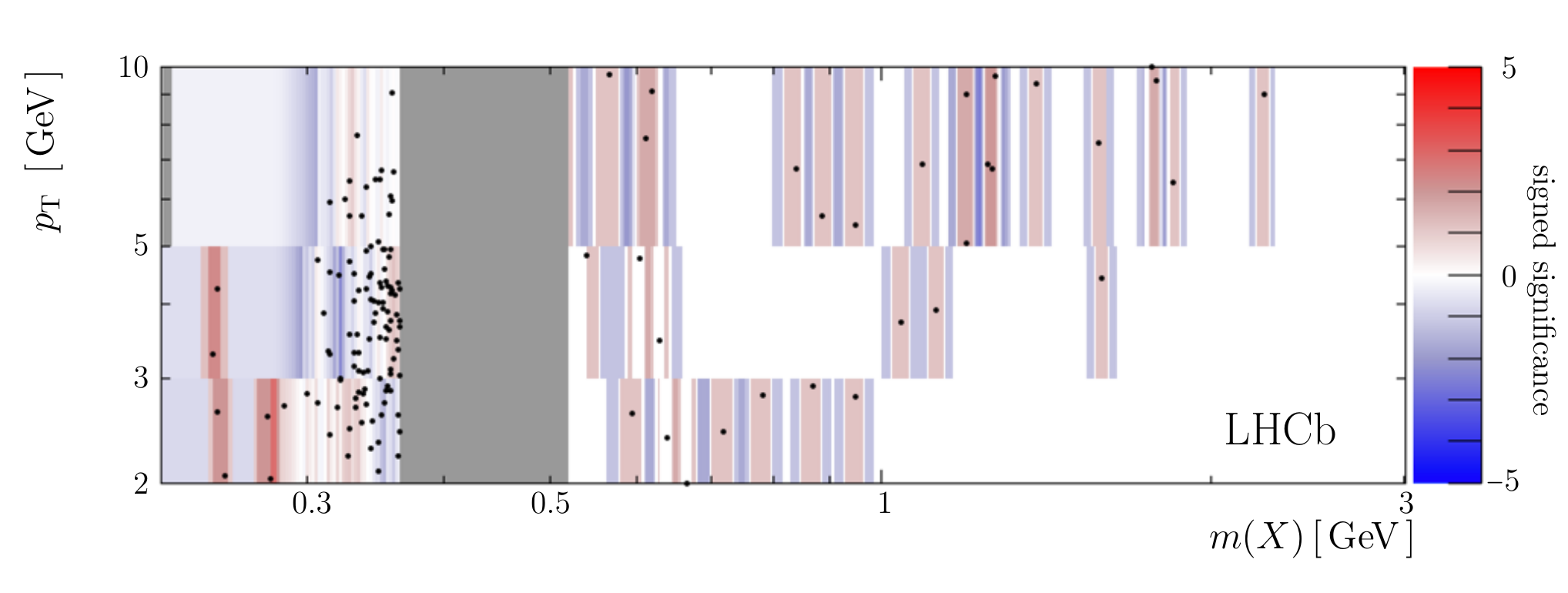} \\\vspace{-0.2cm}
  \includegraphics[width=0.9\textwidth]{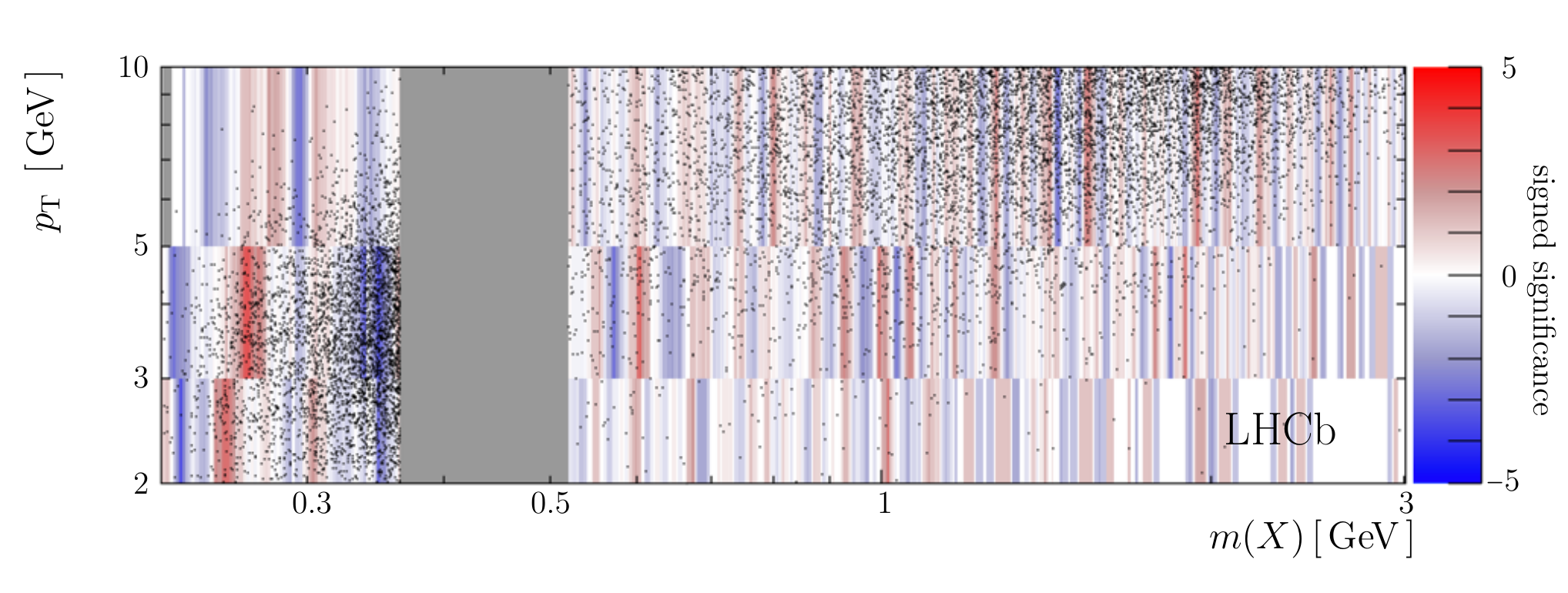}\vspace{-0.2cm}
  \caption{
  Signed local significances for the (top) promptly produced and (bottom) inclusive searches for displaced \xtomm decays. The black points show the individual candidates.
  }
  \label{fig:displ-sgnf}
\end{figure}

Figure~\ref{fig:displ-sgnf} shows the signed local significances for both searches for displaced \xtomm decays.
The largest local excess in the search for a promptly produced long-lived \xboson boson is $2.8\sigma$, which occurs at 280\mev in the $2 < \pt(X) < 3\gev$ bin.
The largest local excess in the inclusive search for displaced \xtomm decays is $3.1\sigma$ at $604\mev$ in the $3 < \pt(X) < 5\gev$ bin.
Both of these correspond to global excesses below $1\sigma$;
therefore, no significant excess is found in either search for displaced \xtomm decays.

\section{Efficiency and luminosity}
\label{sec:eff}

The \xtomm yields are corrected for detection efficiency, which is determined as the product of the trigger, reconstruction, and selection efficiencies.
The hardware trigger efficiency is measured as a function of $\sqrt{\pt(\mu^+)\pt(\mu^-)}$ using a displaced \jpsi calibration sample.
Events selected by the hardware trigger independently of the \jpsi candidate, {\em e.g.}\ due to the presence of a high-\pt hadron, are used to determine the trigger efficiency directly from the data.
The muon reconstruction efficiency is obtained from simulation in bins of $[p(\mu),\eta(\mu)]$.
Scale factors that correct for discrepancies between the data and simulation are determined using a data-driven tag-and-probe approach on an independent sample of $\jpsi\to\mu^+\mu^-$ decays~\cite{LHCb-DP-2013-002}.
The contribution to the selection efficiency from the muon-identification performance is measured in bins of $[\pt(\mu),\eta(\mu)]$ using a highly pure calibration sample of $\jpsi\to\mu^+\mu^-$ decays.
Finally, the contributions from the vertex-quality and prompt-decay criteria are determined from simulation, and validated using a calibration sample of prompt QCD resonance decays to the $\mu^+\mu^-$ final state.

The uncertainty due to the methods  used to determine each of these components of the total efficiency is assessed by
repeating the data-based efficiency studies
on simulated events, where
the difference between the true and efficiency-corrected yields in kinematic bins is used to determine the systematic uncertainty.
These uncertainties are in the 2--5\% range, depending on \xboson-boson kinematics.
Additional uncertainties arise due to the unknown production mechanisms of the \xboson bosons.
The muon reconstruction and identification efficiencies depend on the charged-particle multiplicity.
The corresponding systematic uncertainty
is determined to be 5\%, which covers both minimal and maximal charged-particle multiplicities defined in Table~\ref{tab:fid} at the $2\sigma$ level.
The unknown kinematic distributions in both \pt and $\eta$ within the wide \pt bins used in the analysis lead to sizable uncertainties.
The variation in the efficiencies across the kinematic regions allowed in each bin are used to determine bin-dependent uncertainties that vary from 10 to 30\%.

The $X+b$ analysis uses the SV-based $b$-tagging method described in detail in Ref.~\cite{LHCb-PAPER-2015-016}, though without placing any criteria on the boosted decision tree algorithms; only the presence of an SV is required.
The $b$-tagging efficiency is estimated to be $(65\pm7)\%$, where the uncertainty covers both the variation of the $b$-tagging efficiency across the $b$-jet fiducial region and possible data-simulation discrepancies.
An additional uncertainty arises since the efficiency for a $b$-tagged jet in the fiducial region to be reconstructed with $\pt > 20\gev$  depends on the unknown underlying jet \pt spectrum.
The detector response to jets is studied using the \pt-balance distribution of $\pt({\rm jet})/\pt(Z)$ in nearly back-to-back $Z$-boson$+$jet events using the same data-driven technique as in Ref.~\cite{LHCb-PAPER-2013-058}.
Based on this study, and considering jet \pt spectra as soft as QCD di-$b$-jet production and hard enough to result in negligible inefficiency, this efficiency is estimated to be $(90 \pm 5)\%$.

The searches for displaced \xtomm decays must also account for effects that arise due to the displacement of the SV from the PV.
The relative efficiency of displaced compared to prompt \xtomm decays is obtained as a function of \mx and $\pt(X)$ by resampling prompt \xtomm decay candidates as displaced \xtomm decays, where all displacement-dependent properties
are recalculated based on the resampled SV locations.
The high-precision  material map produced  in Ref.~\cite{LHCb-DP-2018-002} forms the basis of the material-interaction criterion applied in the selection. This map is used to determine where each muon would hit active sensors, and thus, have recorded hits in the VELO.
The resolution on the vertex location and other displacement-dependent properties varies strongly with the location of the first VELO hit on each muon track, though this dependence is largely geometric, making rescaling the resolution of prompt tracks straightforward.
This approach is validated using simulation, where prompt \xtomm decays are used to predict the properties of long-lived \xboson bosons;
these
predictions are found to agree within $2\%$ with the actual values.
The efficiencies at both short and long distances, which  are driven by the muon displacement criterion and the minimum number of VELO hits required to form a track, respectively, are well described.
The dominant uncertainty, which arises due to limited knowledge of how radiation damage has affected the VELO performance,
is estimated to be 5\% by rerunning the resampling method under different radiation-damage hypotheses.

The efficiency of the material-interaction criterion is validated separately using two control samples.
The predicted efficiency for an \xboson boson with the same mass and lifetime as the \KS meson is compared to the efficiency observed in a control sample of \KS decays.
The predicted and observed efficiencies agree to 1\%.
Additionally, in Ref.~\cite{LHCb-DP-2018-002} the expected performance of the material-interaction criterion was shown to agree with the performance observed in a control sample of photon conversions to the $\mathcal{O}(10^{-4})$ level.
Finally, the distribution of the SV locations is unknown, which leads to a 10\% uncertainty in the efficiency determined by comparing the efficiency of an \xboson boson that rarely survives long enough to enter the decay fiducial region to an extremely long-lived \xboson boson.

Most of the data used in this analysis is from data-taking periods that do not yet have fully calibrated luminosities.
Therefore, the efficiency-corrected yield of ${Z/\gamma^* \to \mu^+\mu^-}$ decays observed in the data sample---and the corresponding high-precision LHCb cross-section measurement made using 2015 data~\cite{LHCb-PAPER-2016-021}---are used to infer the luminosity.
A small correction factor is obtained from \pythia~8 to account for the different fiducial regions.
This luminosity determination is validated by also determining the $\Upsilonres(1S)$ differential cross section from this data sample and comparing the results to those published by LHCb using the 2015 data sample~\cite{LHCb-PAPER-2018-002}.
The different fiducial region is again corrected for using a scale factor obtained from \pythia~8.
The results are found to agree to $\approx 5\%$ in each \pt bin, which is assigned as a systematic uncertainty and combined with the 4\% luminosity uncertainty from Ref.~\cite{LHCb-PAPER-2016-021} to obtain the total uncertainty on the luminosity of this data sample.
Based on both of these studies, the luminosity is determined to be $5.1\pm0.3\invfb$.
The minimal dark-photon search~\cite{LHCb-PAPER-2019-031}, which used the same data sample but did not require knowledge of the luminosity, quotes an uncalibrated luminosity value that is 7\% larger.
The efficiency corrections used to infer the luminosity are highly correlated to those used to correct the observed \xtomm yields, which is accounted for when determining the total normalization uncertainties.

\section{Cross-section results}
\label{sec:results}

\begin{figure}[t!]
  \centering
  \includegraphics[width=0.9\textwidth]{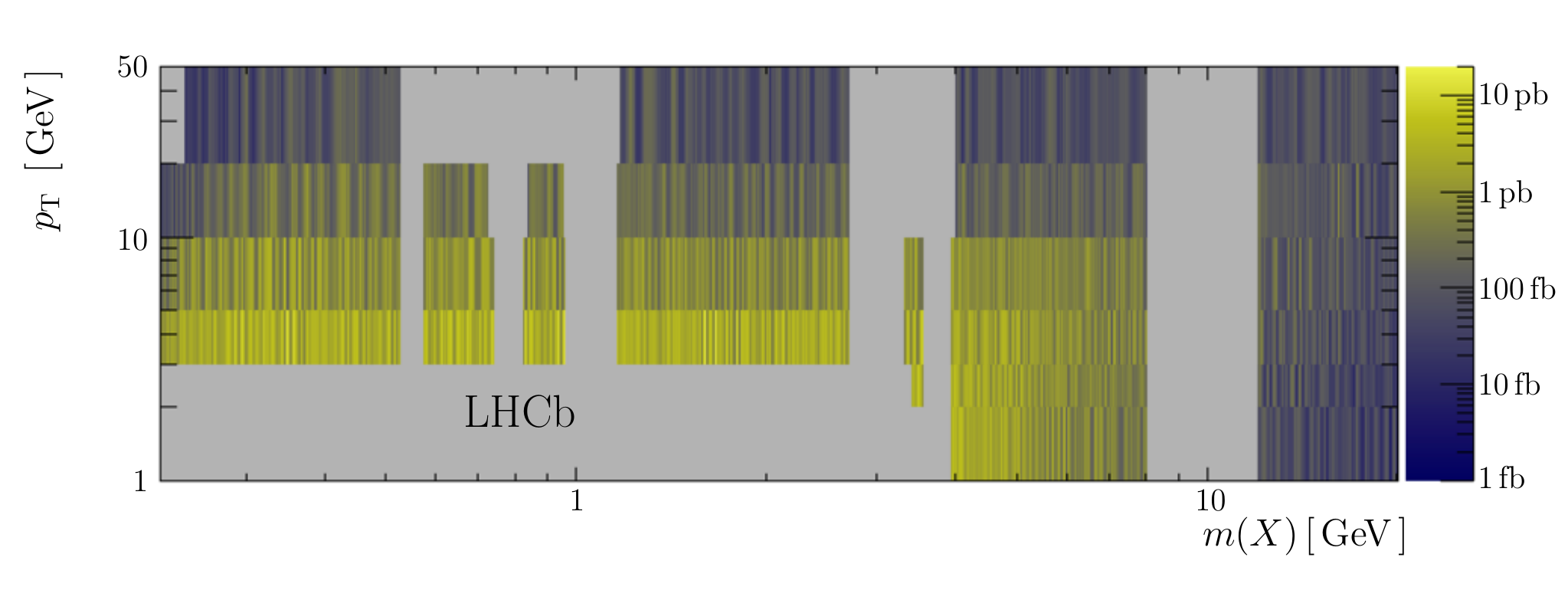}\\\vspace{-0.2cm}
  \includegraphics[width=0.9\textwidth]{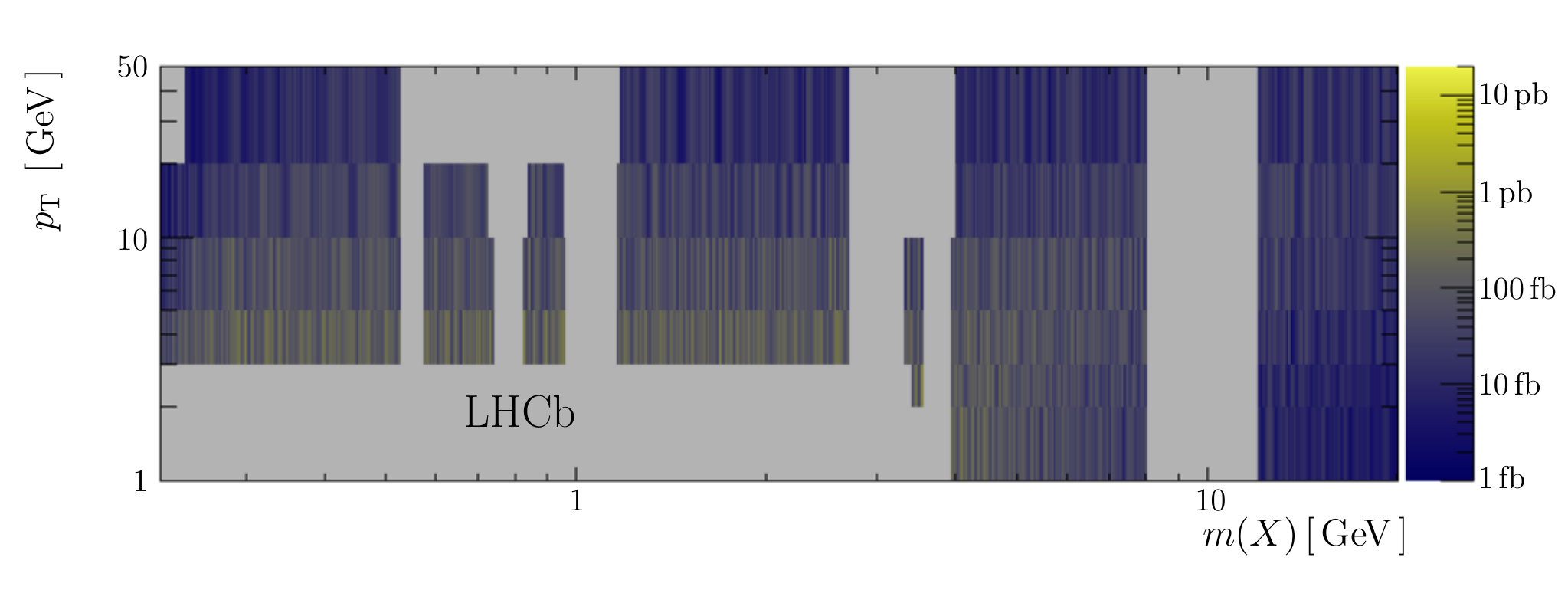}\vspace{-0.2cm}
  \caption{
  Upper limits at 90\% confidence level on the cross section $\sigma(\xtomm)$  in the $\mx < 20\gev$ region for the (top) inclusive and (bottom) associated beauty searches for prompt \xtomm decays.
  }
  \label{fig:prompt-lims}
\end{figure}

\begin{figure}[t!]
  \centering
  \includegraphics[width=0.9\textwidth]{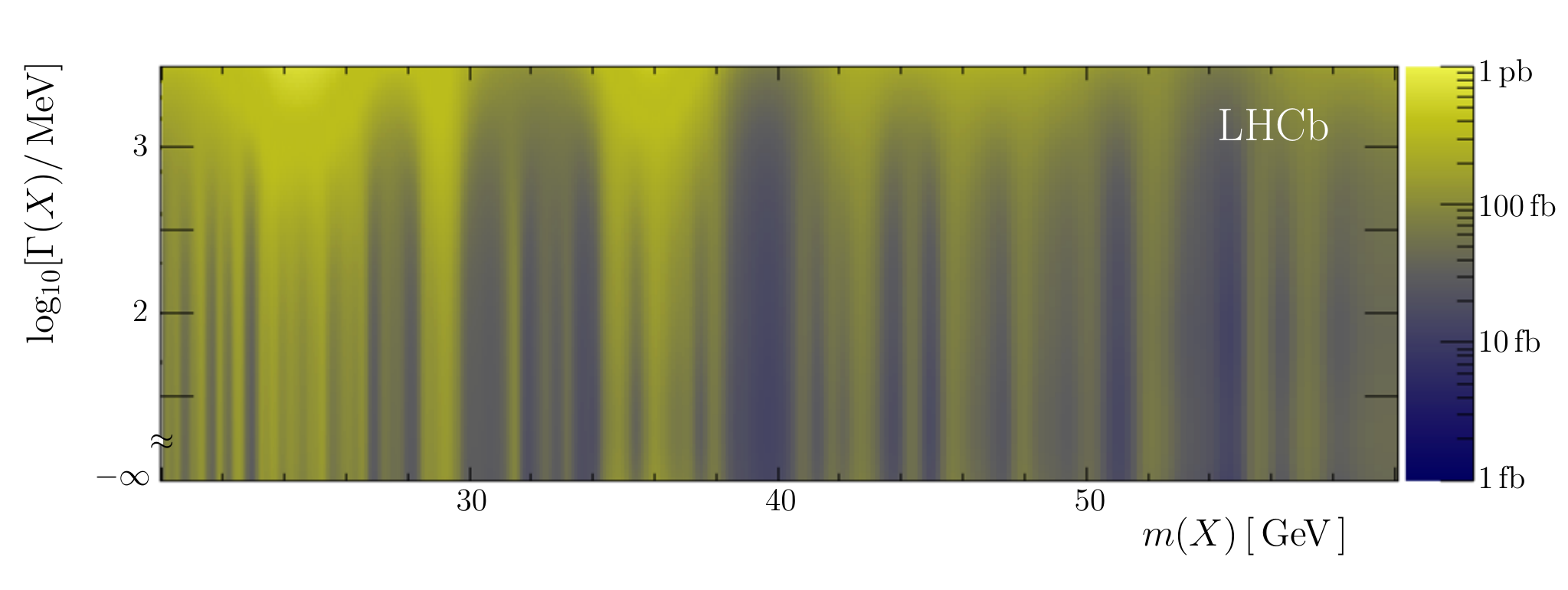} \\ \vspace{-0.2cm}
  \includegraphics[width=0.9\textwidth]{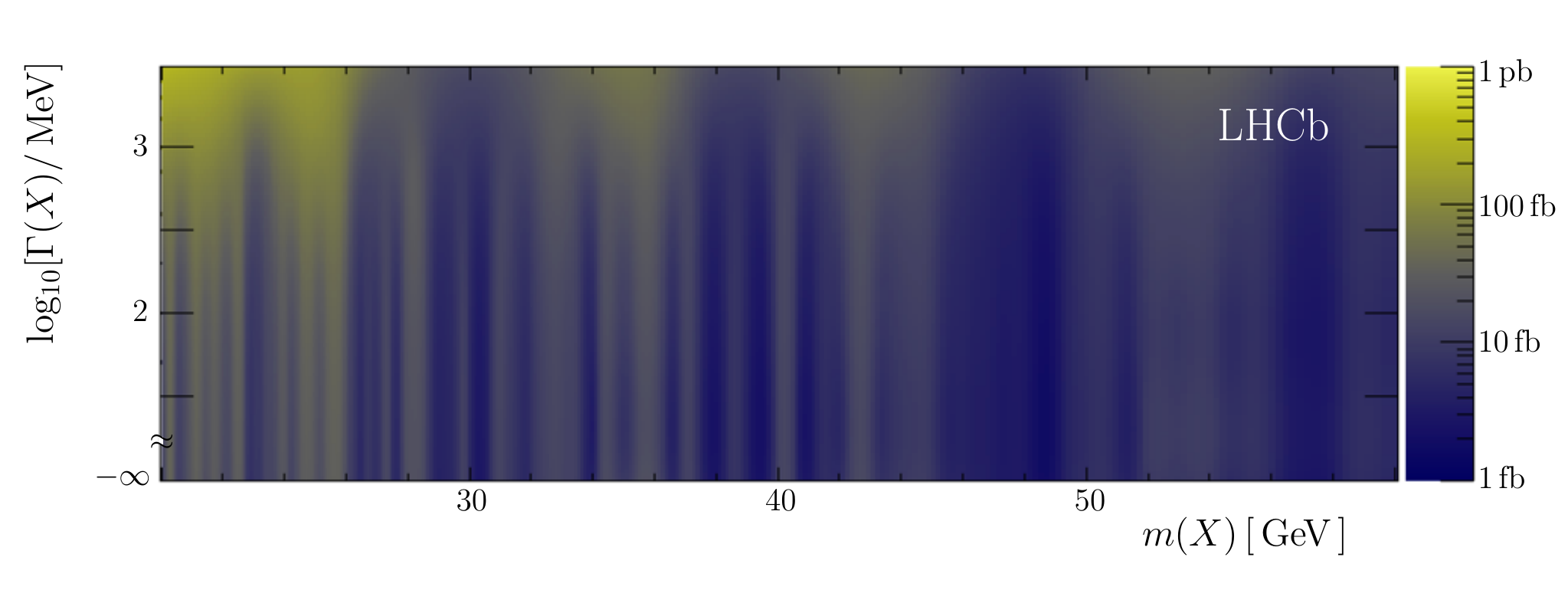}\vspace{-0.2cm}
  \caption{
  Upper limits at 90\% confidence level on the cross section $\sigma(\xtomm)$  in the $\mx > 20\gev$ region for the (top) inclusive and (bottom) associated beauty searches for prompt \xtomm decays.
  }
  \label{fig:prompt-lims-highmass}
\end{figure}

\begin{figure}[t!]
  \centering
  \includegraphics[width=0.9\textwidth]{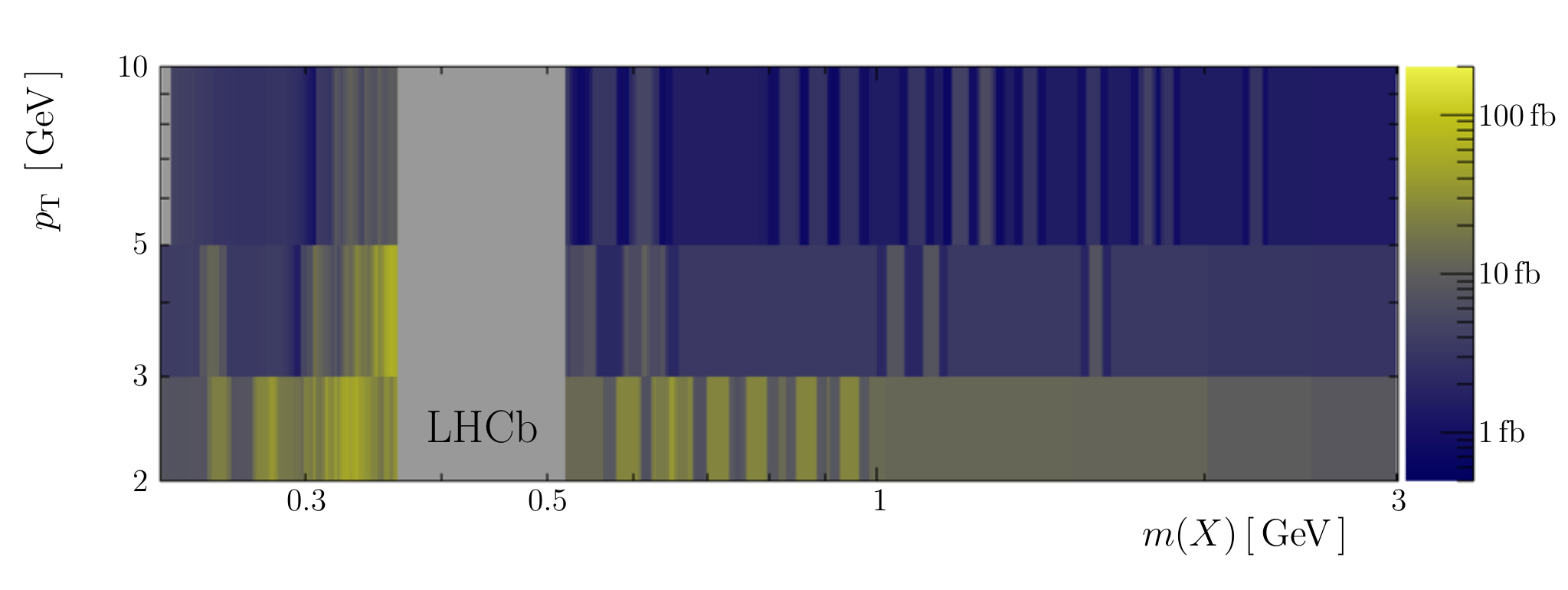}\\\vspace{-0.2cm}
  \includegraphics[width=0.9\textwidth]{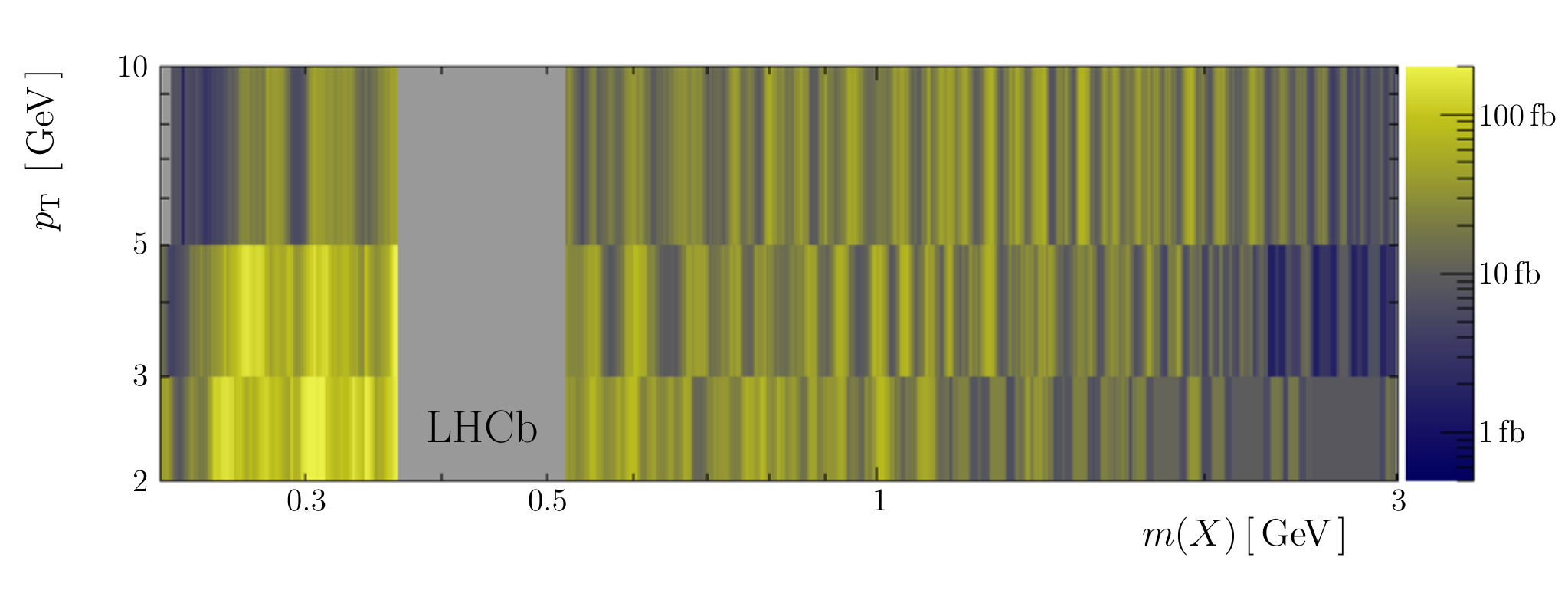}\vspace{-0.2cm}
  \caption{
  Upper limits at 90\% confidence level on the cross section $\sigma(\xtomm)$ for the (top)~promptly produced and (bottom) inclusive searches for displaced \xtomm decays.
  }
  \label{fig:displ-lims}
\end{figure}

The upper limits on the signal yields obtained in Sec.~\ref{sec:search} are normalized using the efficiencies and luminosity described in Sec.~\ref{sec:eff}.
The systematic uncertainties on the signal yield, efficiency, and luminosity are included in the profile likelihood when determining the cross-section upper limits.
These uncertainties are described in detail in Secs.~\ref{sec:search} and \ref{sec:eff}, and summarized in Table~\ref{tab:syst}.
The resulting upper limits at 90\% confidence level on $\sigma(\xtomm)$ for all searches are shown in Figs.~\ref{fig:prompt-lims}--\ref{fig:displ-lims}, and provided numerically in Ref.~\cite{Supp}.

\begin{table}[t!]
    \caption{\label{tab:syst} Summary of systematic uncertainties. The luminosity and efficiency uncertainties are highly correlated, which is accounted for when obtaining the total uncertainties. }\vspace{-0.5cm}
    \begin{center}
      \begin{tabular}{lc}
        \toprule
        Source & Relative uncertainty \\
        \midrule
        Signal model & 5\% \\
        Background model & data driven, see Sec.~\ref{sec:search} \\
        \midrule
        Trigger, reconstruction, selection & 2--5\% (bin dependent) \\
        Charged-particle multiplicity & 5\% \\
        \xboson kinematics & 10--30\% (bin dependent) \\
        $b$-jet selection & 11\% ($X+b$ only) \\
        SV selection & 5\% (SV-based only) \\
        \xboson SV distribution & 10\% (SV-based only) \\
        \midrule
        Luminosity & 6\% \\
        \midrule
        Total & 11--30\% (bin dependent) \\
        \bottomrule
      \end{tabular}
  \end{center}
\end{table}

\begin{figure}[t!]
  \centering
  \includegraphics[width=0.9\textwidth]{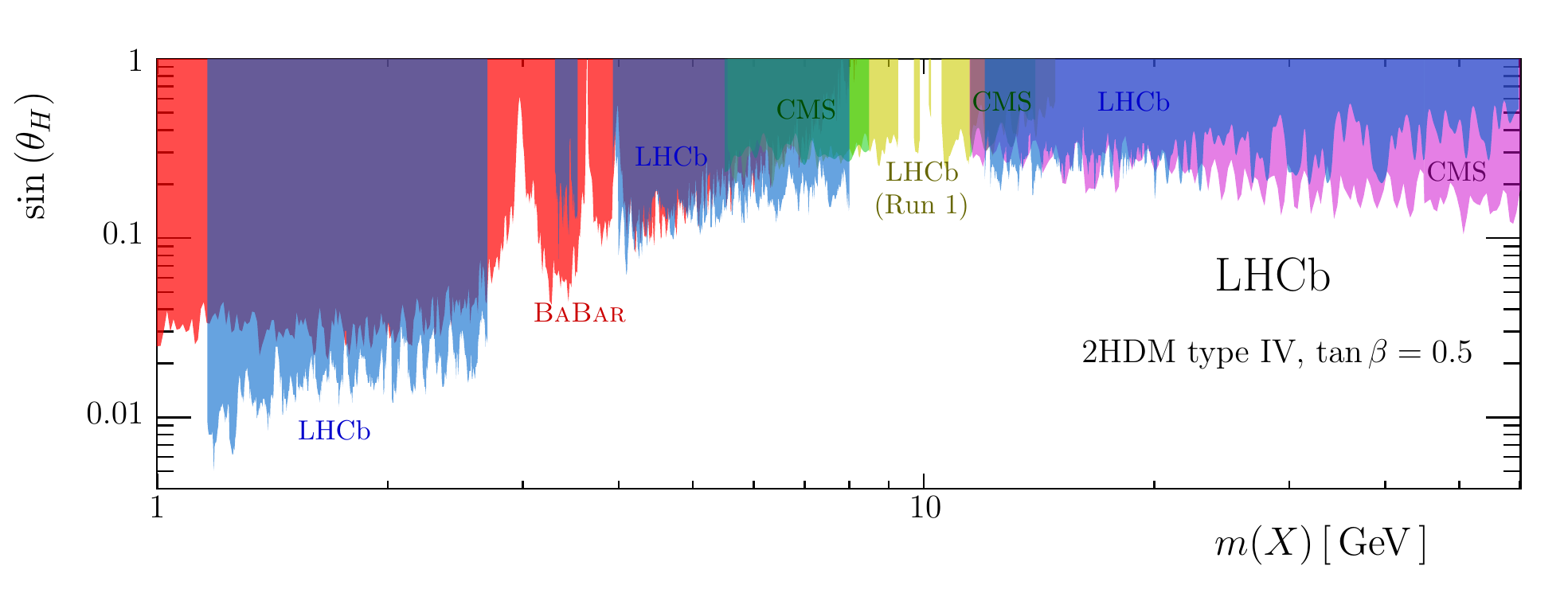}\vspace{-0.2cm}
  \caption{
  Upper limits at 90\% confidence level on the $X$--$H$ mixing angle, $\theta_H$, for the 2HDM scenario discussed in the text (blue) from this analysis compared with existing limits from (red) {\sc BaBar}~\cite{Lees:2012iw}, (green)~CMS Run~1~\cite{Chatrchyan:2012am}, (magenta) CMS Run~2~\cite{Sirunyan:2019wqq} and (yellow) LHCb Run~1~\cite{LHCb-PAPER-2018-008}.
  }
  \label{fig:2hdm}
\end{figure}

The model-independent limits in Figs.~\ref{fig:prompt-lims}--\ref{fig:prompt-lims-highmass} can be used to place constraints on any model that would produce a promptly decaying low-mass dimuon resonance within the fiducial region of Table~\ref{tab:fid}.
For example, models where a complex scalar singlet is added to the two-Higgs doublet (2HDM) potential often feature a light pseudoscalar boson that can decay into the dimuon final state; see, {\em e.g.}, Ref.~\cite{Branco:2011iw}.
References~\cite{Haisch:2016hzu,Haisch:2018kqx} considered the scenario where the pseudoscalar boson acquires all of its couplings to SM fermions through its mixing with the Higgs doublets;
the corresponding $X$--$H$ mixing angle is denoted as $\theta_{H}$.
Figure~\ref{fig:2hdm} shows that world-leading constraints are placed on $\theta_{H}$ by the $\sigma(\xtomm)$ limits shown in Figs.~\ref{fig:prompt-lims}--\ref{fig:prompt-lims-highmass}; these constraints are twice as strong in the $\mathcal{O}(\gev)$ region as those obtained by recasting the dark-photon results in Ref.~\cite{LHCb-PAPER-2019-031}.
Furthermore, assuming the $X+b\bar{b}$ topology produced by this type of model permits direct comparison with the excess seen by CMS in this final state~\cite{Sirunyan:2018wim}.
For this scenario, the $X+b$ limits from Fig.~\ref{fig:prompt-lims-highmass} are about 20 times lower than the excess observed by CMS.

\begin{figure}[t!]
  \centering
  \includegraphics[width=0.9\textwidth]{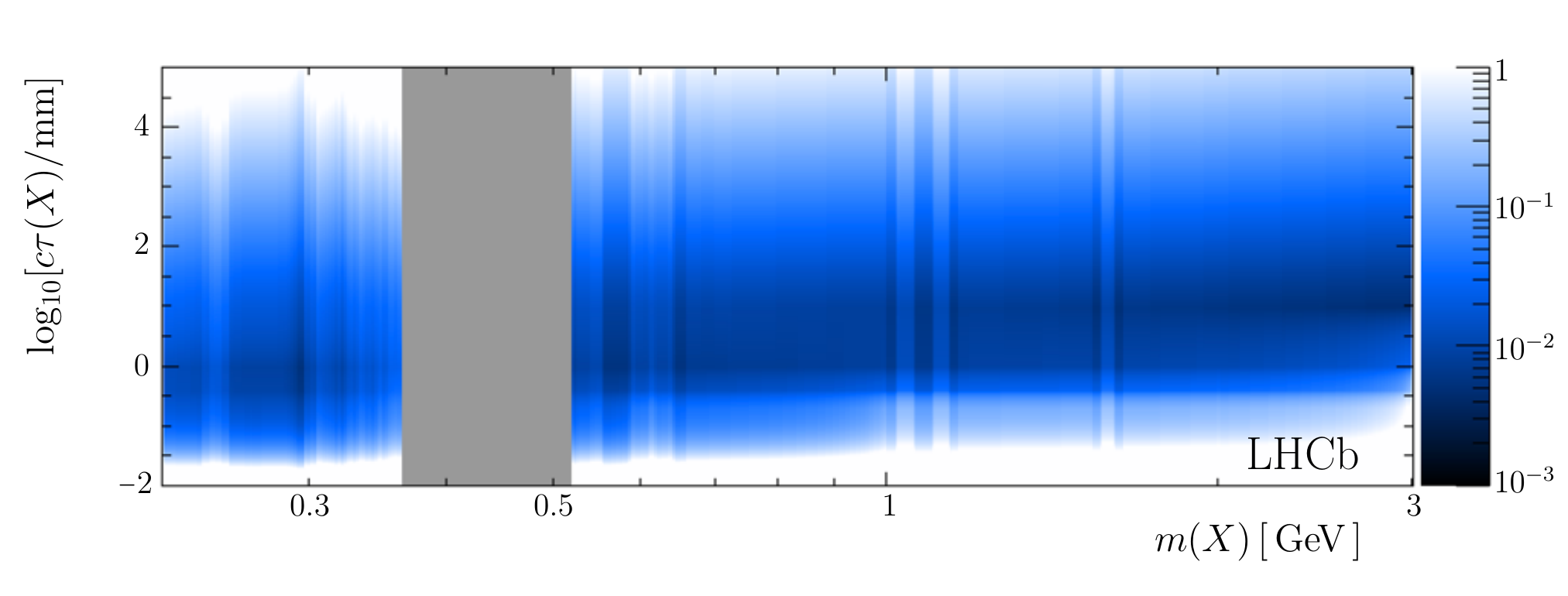}\vspace{-0.2cm}
  \caption{
  Upper limits at 90\% confidence level on the $\gamma$--$Z_{\rm HV}$ kinetic mixing strength for the HV scenario discussed in the text. Here, $X$ denotes a composite HV vector boson.
  }
  \label{fig:HV}
\end{figure}

The limits on displaced \xtomm decays in Fig.~\ref{fig:displ-lims} can also be used to place constraints on specific models.
One example is HV scenarios that exhibit confinement, which result in a large multiplicity of light hidden hadrons from showering processes~\cite{Pierce:2017taw}.
These hidden hadrons typically have low \pt and decay displaced from the proton-proton collision.
Figure~\ref{fig:HV} shows the limits placed on this type of HV scenario by the search for displaced \xtomm decays.
These are the most stringent constraints to date.
Specifically, constraints are placed on the kinetic-mixing strength between the photon and a heavy HV boson, $Z_{\rm HV}$, with photon-like couplings. The kinematics of the hidden hadrons depend upon the average HV hadron multiplicity, $\langle N_\mathrm{HV}\rangle$, and are largely independent of the model parameter space. In Fig.~\ref{fig:HV} $\langle N_\mathrm{HV}\rangle$ is fixed at $\approx 10$ for all hidden hadron masses.
These are the first results that constrain the kinetic-mixing strength to be less than unity in this mass region.

\section{Summary}
\label{sec:sum}

In summary, searches are performed for low-mass dimuon resonances produced in proton-proton collisions at a center-of-mass energy of 13\tev
using a data sample corresponding to an integrated luminosity of 5.1\invfb collected with the LHCb detector.
The \xtomm decays can be either prompt or displaced from the proton-proton collision, where in both cases the requirements placed on the event and the assumptions made about the production mechanisms are kept as minimal as possible.
Two variations are performed of the search for prompt \xtomm decays:
an inclusive version,
and one where the \xboson boson is required to be produced in association with a beauty quark.
Two variations are also considered of the search for displaced \xtomm decays:
an inclusive version,
and one where the \xboson boson is  required  to be produced promptly in the proton-proton collision.
The searches for prompt \xtomm decays explore the mass range from near the dimuon threshold up to 60\gev, with nonnegligible \xboson widths considered above 20\gev.
The searches for displaced \xtomm decays consider masses up to 3\gev.
None of the searches finds evidence for a signal, and 90\% confidence-level exclusion limits are placed on the \xtomm cross sections, each with minimal model dependence.

\section*{Acknowledgements}
%
%
\noindent We express our gratitude to our colleagues in the CERN
accelerator departments for the excellent performance of the LHC. We
thank the technical and administrative staff at the LHCb
institutes.
We acknowledge support from CERN and from the national agencies:
CAPES, CNPq, FAPERJ and FINEP (Brazil); 
MOST and NSFC (China); 
CNRS/IN2P3 (France); 
BMBF, DFG and MPG (Germany); 
INFN (Italy); 
NWO (Netherlands); 
MNiSW and NCN (Poland); 
MEN/IFA (Romania); 
MSHE (Russia); 
MinECo (Spain); 
SNSF and SER (Switzerland); 
NASU (Ukraine); 
STFC (United Kingdom); 
DOE NP and NSF (USA).
We acknowledge the computing resources that are provided by CERN, IN2P3
(France), KIT and DESY (Germany), INFN (Italy), SURF (Netherlands),
PIC (Spain), GridPP (United Kingdom), RRCKI and Yandex
LLC (Russia), CSCS (Switzerland), IFIN-HH (Romania), CBPF (Brazil),
PL-GRID (Poland) and OSC (USA).
We are indebted to the communities behind the multiple open-source
software packages on which we depend.
Individual groups or members have received support from
AvH Foundation (Germany);
EPLANET, Marie Sk\l{}odowska-Curie Actions and ERC (European Union);
A*MIDEX, ANR, Labex P2IO and OCEVU, and R\'{e}gion Auvergne-Rh\^{o}ne-Alpes (France);
Key Research Program of Frontier Sciences of CAS, CAS PIFI, and the Thousand Talents Program (China);
RFBR, RSF and Yandex LLC (Russia);
GVA, XuntaGal and GENCAT (Spain);
the Royal Society
and the Leverhulme Trust (United Kingdom).

\addcontentsline{toc}{section}{References}
\bibliographystyle{LHCb}
\bibliography{main,standard,LHCb-PAPER,LHCb-CONF,LHCb-DP,LHCb-TDR,dark-photons}

\newpage
\centerline
{\large\bf LHCb collaboration}
\begin
{flushleft}
\small
R.~Aaij$^{31}$,
C.~Abell{\'a}n~Beteta$^{49}$,
T.~Ackernley$^{59}$,
B.~Adeva$^{45}$,
M.~Adinolfi$^{53}$,
H.~Afsharnia$^{9}$,
C.A.~Aidala$^{82}$,
S.~Aiola$^{25}$,
Z.~Ajaltouni$^{9}$,
S.~Akar$^{64}$
% ...

{\footnotesize \it

% ...
$ ^{9}$Universit{\'e} Clermont Auvergne, CNRS/IN2P3, LPC, Clermont-Ferrand, France\\
% ...
$ ^{25}$INFN Sezione di Milano, Milano, Italy\\
% ...
$ ^{31}$Nikhef National Institute for Subatomic Physics, Amsterdam, Netherlands\\
% ...
$ ^{45}$Instituto Galego de F{\'\i}sica de Altas Enerx{\'\i}as (IGFAE), Universidade de Santiago de Compostela, Santiago de Compostela, Spain\\
% ...
$ ^{49}$Physik-Institut, Universit{\"a}t Z{\"u}rich, Z{\"u}rich, Switzerland\\
% ...
$ ^{53}$H.H. Wills Physics Laboratory, University of Bristol, Bristol, United Kingdom\\
% ...
$ ^{59}$Oliver Lodge Laboratory, University of Liverpool, Liverpool, United Kingdom\\
% ...
$ ^{64}$University of Cincinnati, Cincinnati, OH, United States\\
% ...
$ ^{67}$Syracuse University, Syracuse, NY, United States\\
% ...
$ ^{82}$University of Michigan, Ann Arbor, United States, associated to $^{67}$\\
% ...
\medskip
}
\end{flushleft}

\end{document}